\documentclass[reprint,
superscriptaddress,
frontmatterverbose, 
showpacs,
preprintnumbers,
nofootinbib,
nobibnotes,
 amsmath,
 amssymb, 
 aps,
 longbibliography,
]{revtex4-1}%

\usepackage{scalerel}
\usepackage{dcolumn}
\usepackage{bm}
\usepackage[utf8]{inputenc}
\usepackage{amsmath}
\usepackage[dvipsnames]{xcolor}
\usepackage{graphicx}
\usepackage{fancyhdr}
\usepackage{lastpage}
\usepackage{graphicx}
\usepackage{amsmath}
\usepackage[T1]{fontenc}
\usepackage{physics,slashed}
\usepackage{amssymb}
\usepackage{mathtools}
\usepackage[colorlinks=true,linkcolor=blue,citecolor=blue]{hyperref}
\usepackage[capitalise]{cleveref}
\usepackage{multirow}
\usepackage[normalem]{ulem}

\newcommand{\reffig}[1]{Fig.~\ref{#1}}

\newcommand{\refapp}[1]{Appendix~\ref{#1}}
\newcommand{\reftab}[1]{Table~\ref{#1}}
\newcommand{\refref}[1]{Ref.~\cite{#1}}

\def\eg{\emph{e.g.}}

\renewcommand{\phi}{\varphi}
\renewcommand{\epsilon}{\varepsilon}
\def\Z2{\text{Z}_2}

\newcommand\wh[1]{\hstretch{2}{\hat{\hstretch{.5}{#1}}}}
\newcommand{\makevec}{\wh}

\newcounter{CommentCount}
\usepackage{fontawesome} 
\newcommand{\orcid}[1]{\href{https://orcid.org/#1}{#1}}

\begin{document}

\preprint{FTPI-MINN-20-25,
IPPP/20/32}

\title{A Dark Seesaw Solution to Low Energy Anomalies:
\\MiniBooNE, the muon $(g-2)$, and BaBar}

\author{Asli Abdullahi}
\email{asli.abdullahi@durham.ac.uk}
\thanks{\orcid{0000-0002-6122-4986}}
\affiliation{Institute for Particle Physics Phenomenology, Department of
Physics, Durham University, South Road, Durham DH1 3LE, United Kingdom}
\author{Matheus Hostert}
\email{mhostert@umn.edu}
\thanks{\orcid{0000-0002-9584-8877}}
\affiliation{School of Physics and Astronomy, University of Minnesota, Minneapolis, MN 55455, USA}
\affiliation{William I. Fine Theoretical Physics Institute, School of Physics and Astronomy, University of
Minnesota, Minneapolis, MN 55455, USA}
\affiliation{Perimeter Institute for Theoretical Physics, Waterloo, ON N2J 2W9, Canada}
\author{Silvia Pascoli}
\email{silvia.pascoli@durham.ac.uk}
\thanks{\orcid{0000-0002-2958-456X}}
\affiliation{Institute for Particle Physics Phenomenology, Department of
Physics, Durham University, South Road, Durham DH1 3LE, United Kingdom}

\date{\today}

\begin{abstract}
A recent update from MiniBooNE has strengthened the observed $4.8\sigma$ excess of $e$-like events. Motivated by this and other notable deviations from standard model predictions, such as the  muon $(g-2)$, we propose a solution to low energy anomalies through a dark neutrino sector. The model is renormalizable and can also explain light neutrino masses with an anomaly-free and dark $U(1)^\prime$ gauge symmetry broken at the GeV scale. Large kinetic mixing leads to s-channel production of heavy neutral leptons at $e^+e^-$ colliders, where we point out and explain a $\gtrsim 2\sigma$ excess observed in the BaBar monophoton data. Our model is also compatible with anomalous $e$-like events seen at old accelerator experiments, as well as with an excess of double vertex signatures observed at CCFR. 
\end{abstract}

\maketitle

\section{Introduction}
 The discovery of neutrino oscillations~\cite{Fukuda:1998ah,Ahmad:2002jz,Eguchi:2002dm}, and consequently of neutrino masses and mixing, implies that the Standard Model (SM) of particle physics is incomplete. Many extensions have been proposed to explain the origin of neutrino masses, with the Type-I seesaw mechanism~\cite{Minkowski:1977sc,Mohapatra:1979ia,GellMann:1980vs,Yanagida:1979as,Lazarides:1980nt,Mohapatra:1980yp,Schechter:1980gr,Cheng:1980qt,Foot:1988aq} and its variants being the most well studied.
 Heavy neutral leptons (HNL) are the hallmark of such models and carry a lepton number violating (LNV) Majorana mass, which, barring theoretical prejudice, could take any value from sub-eV to $10^{16}$ GeV. In recent years, renewed attention has been devoted to the MeV - GeV mass scale, as such states can be searched for in an expanding program of fixed-target, meson decay, and collider experiments~\cite{Atre:2009rg,Essig:2013lka,Drewes:2015iva,Bondarenko:2018ptm,Beacham:2019nyx,Ballett:2019bgd,Berryman:2019dme}, having consequences for cosmology and the baryon asymmetry of the Universe~\cite{Fukugita:1986hr,Davidson:2008bu}. Two approaches are typically adopted: one of minimality, in which only new neutral fermions are introduced, e.g.~\cite{Asaka:2005pn}, and, more recently, one in which the HNLs are considered part of a richer low energy dark sector~\cite{Pospelov:2011ha,Harnik:2012ni,Batell:2016zod,Farzan:2016wym,DeRomeri:2017oxa,Magill:2018jla,Bertuzzo:2018ftf,Bertuzzo:2018itn,Ballett:2019cqp,Ballett:2019pyw,Coloma:2019qqj,Fischer:2019fbw,Cline:2020mdt,Berbig:2020wve}, all the more compelling in view of the large abundance of dark matter in our Universe~\cite{Boehm:2003hm,Boehm:2003ha,Pospelov:2007mp,Pospelov:2008zw}. It has been pointed out that in the second approach the phenomenology can have unique features, requiring the reevaluation of existing bounds and offering new signatures, especially in the presence of multiple portals to the SM~\cite{Ballett:2018ynz}. Such an extension of the SM would leave imprints, not just in neutrino experiments, but also in e.g. dark photon and dark scalar searches. Interestingly, some anomalies are present in these areas.

In this \emph{letter}, we propose a coherent explanation of several experimental anomalies, generating the correct scale for the light neutrino masses. We simultaneously explain the excess of $e$-like events observed at MiniBooNE~\cite{Aguilar-Arevalo:2018gpe} and the muon $\Delta a_\mu = (g-2)_\mu$ anomaly~\cite{Bennett:2006fi}. We also point out some less-often discussed anomalies in existing data which are compatible with the predictions of our model. These include a mild excess of monophoton events at BaBar~\cite{Lees:2017lec}, the anomalous $\nu_e$-appearance observed by past accelerator experiments, such as PS-191~\cite{Bernardi:1986hs} and E-816~\cite{Astier:1989vc}, and the double neutral vertex events in CCFR~\cite{Budd:1992vt,deBarbaro:1992hb}. 
We show how these results emerge within a coherent picture and that they are, in fact, highly correlated when interpreted under our hypothesis.
This is achieved within an anomaly-free model of a spontaneously-broken and secluded $U(1)^\prime$ gauge symmetry, providing a concrete model for the phenomenological idea put forward in Ref.~\cite{Ballett:2018ynz}. The presence of sterile and dark vector-like neutrinos leads to light neutrino masses via a generalized inverse seesaw~\cite{Dev:2012sg,Dev:2012bd,LopezPavon:2012zg}, modified by the interactions in the dark sector.

\section{Model}
We extend the SM gauge symmetry with a secluded $U(1)^\prime$, accompanied by a dark~\footnote{In the following, we refer to particles charged under the $U(1)^\prime$ gauge symmetry as ``dark".} complex scalar $\Phi$ with charge $Q_X$ that breaks the symmetry at sub-GeV scales. Generically, our fermionic sector comprises of $d$ vector-like dark neutrinos, $\makevec{\nu}_D = \makevec{\nu}_{D_L}+\makevec{\nu}_{D_R}$, also charged under the $U(1)^\prime$ with charge $Q_X$, guaranteeing anomaly cancellation in each dark neutrino family. A neutrino portal to the SM is then achieved by $n$ completely sterile states, $\makevec{N}$. 

The full Lagrangian is given by
\begin{align}
\mathcal{L} &\supset \mathcal{L}_{\rm SM}
- \frac{1}{4} X_{\mu\nu}X^{\mu\nu} -\frac{\sin \chi}{2} X_{\mu\nu}B^{\mu\nu} 
\\
& + \left(D_\mu \Phi\right)^\dagger \left(D^\mu \Phi\right) - V(\Phi)- \lambda_{\Phi H} \left| H \right|^2\left| \Phi \right|^2 \nonumber\\
&+\overline{\makevec{\nu}_N}i\slashed{\partial}\makevec{\nu}_N+ \overline{\makevec{\nu}_D}i\slashed{D}_X\makevec{\nu}_D - \bigg[ (\overline{L} \widetilde{H}) Y \makevec{\nu}_N^c +\frac{1}{2}\overline{\makevec{\nu}_N}{M_N}\makevec{\nu}_N^c
\nonumber\\   
&+ \overline{\makevec{\nu}_N} \left( Y_L \makevec{\nu}_{D_L}^c\Phi + Y_R \makevec{\nu}_{D_R}\Phi^*\right)  + \overline{\makevec{\nu}_{D}} M_X \makevec{\nu}_{D} + \text{h.c.}\bigg]\,,\nonumber
\end{align}
where flavor indices are implicit, and we write the kinetic mixing between hypercharge and the $U(1)^\prime$ mediator $X_\mu$, as well as scalar mixing between the Higgs and $\Phi$ explicitly. Here, $X_{\mu \nu} \equiv \partial_\mu X_\nu - \partial_\nu X_\mu$, $\slashed{D}_X \equiv \slashed{\partial} - i Q_X g_X \slashed{X}$, and $Q_X[\nu_{D_L}] = Q_X[\nu_{D_R}] = 1$. 
The scalars $\Phi$ and $H$ acquire VEVs, $v_\Phi \simeq \mathcal{O}(500)$~MeV and $v_H \simeq 246$~GeV, respectively. After the electroweak and dark symmetries are spontaneously broken, taking $\makevec{\nu}_f \equiv \left( \begin{matrix} \makevec{\nu}_\alpha^c & \makevec{\nu}_N^c & \makevec{\nu}_{D_L}^c& \makevec{\nu}_{D_R} \end{matrix} \right)^T$, the neutrino mass matrix reads
\begin{equation}\label{eq:mass_matrix}
    \mathcal{L}_{\nu-\text{mass}} = \frac{1}{2}\overline{\makevec{\nu}_f^c} \left( 
    \begin{matrix} 0 & M_D & 0 & 0 \\ 
                   M_D^T & M_N & \Lambda_L & \Lambda_R \\
                   0 & \Lambda_L^T & 0 & M_X \\
                   0 & \Lambda_R^T & M_X^T & 0 \end{matrix} 
    \right) \makevec{\nu}_f + \text{ h.c.}\,,
\end{equation}
where  $M_D \equiv Y v_H / \sqrt{2}$ and $\Lambda_{L,R} \equiv Y_{L,R} \,v_\Phi/\sqrt{2}$. We diagonalize the mass matrix with a unitary matrix $U$, defined in terms of sub-blocks $U \equiv \left( \begin{matrix} U_\alpha & U_N & U_{D_L} & U_{D_R} \end{matrix} \right)^T$, such that $\makevec{\nu}_m = U \,\makevec\nu_f \equiv \left( \begin{matrix} \nu & N \end{matrix} \right)^T$ contains the light neutrinos $\nu$ and the $(n+2d)$ HNLs $N$. At tree-level, the mostly-active neutrinos get a mass as in the inverse~\cite{Mohapatra:1986bd,GonzalezGarcia:1988rw} and extended seesaw~\cite{Barry:2011wb,Zhang:2011vh} models. At the one-loop-level, however, we find an independent finite contribution proportional to $M_N$~\cite{future}. This is the same contribution found in Ref.~\cite{Ballett:2019cqp}, and is analogous to the \emph{minimal radiative} inverse seesaw~\cite{Dev:2012sg,Dev:2012bd,LopezPavon:2012zg}. These independent tree- and loop-level contributions can have opposite signs, leading to cancellations if $M_X \lesssim M_N$. We exploit this fact to achieve neutrino masses compatible with current data. We neglect loop corrections to other mass parameters in the matrix.

\renewcommand{\arraystretch}{1.3}
\begin{table*}[t]
    \centering
    \scalebox{0.97}{
    \begin{tabular}{|c|cccc|c|c|c c c|c c c|c c c c c c| c c c|}
    \hline
    \multirow{2}{*}{BP} & \multirow{2}{*}{MB} & \multirow{2}{*}{$\Delta a_\mu$} & \multirow{2}{*}{BB}& \multirow{2}{*}{Acc} &  \multirow{2}{*}{\,$\alpha_D$\,}  & $m_3$ & $m_4$& $m_5$ & $m_6$ & $|V_{43}|^2$ & $|V_{53}|^2$ & $|V_{63}|^2$ & \multicolumn{6}{c|}{$\mathcal{B}(Z^\prime \to N_j N_k)/\%$} & \multicolumn{3}{c|}{$c\tau^0$/cm}
    \\
     & & & & &  &\, /eV \,& \multicolumn{3}{c|}{/MeV} & \multicolumn{3}{c|}{$/10^{-8}$} & 44 & 45 & 46 & 55 & 56 & 66 & $N_4$ & $N_5$ & $N_6$ \\
    \hline \hline
    {A} & \checkmark
    & \checkmark & \checkmark
    & (\checkmark) & $0.39$  & $0.05$ & $35$ & $120$ & $185$ & $0$ & $22.2$ & $0$ & $0$ & $5.4$ & $0$ & $0$ & $95$ & $0$ & $1.6 \times 10^{13}$ & $3.0$ & $0.26$ \\
     {B} & \checkmark & \checkmark & \checkmark & \checkmark & $0.32$  & $0.05$ & $74$ & $146$ & $220$ & $13.6$ & $26.5$ & $123$ & $0.15$ & $11$ & $0.48$ & $1.6$ & $86$ & $0.59$ & $1.1 \times 10^7$ & $2.2$  & $0.14$\\
    \hline
    \end{tabular}}
    \caption{Benchmark points used in this study, where $m_{Z^\prime} = 1.25$ GeV and $m_{\phi^\prime} = 1.6$~GeV always. Here, the $V_{ij} \equiv U_{D_L i}^*U_{D_L j} - U_{D_R i}^*U_{D_R j}$ are the mixing factors in $Z^\prime N_i \nu_j$ vertices, and $\alpha_D = g_X^2 /4\pi$. Note that $Z^\prime \to \nu_3 \nu_3$ is negligible for the mixings considered. We refer to the MiniBooNE excess as MB, the BaBar excess as BB, and the accelerator experiments as Acc. The zeroes in BP-A are protected by a left-right symmetry ($\Lambda_L = \Lambda_R$).}
    \label{tab:BPmain}
\end{table*}

The massive dark photon, scalar, and HNLs only couple to the SM via portal operators.
After symmetry breaking, the model has two CP-even scalars, the SM Higgs $h^\prime$, which contains a small $\Phi$ component with scalar mixing $\theta \simeq (\lambda_{\Phi H}/ 2 \lambda_H)\times (v_\Phi / v_{H})$, where $\lambda_H$ is the quartic coupling of the Higgs, and a light mostly-dark $\phi^\prime$.
In the neutral gauge boson mass basis, we have a light $Z^\prime$ vector boson that couples predominantly to the dark sector current ($J_D^\mu$), as well as to the SM electromagnetic (EM), and neutral current (NC),
\begin{equation}
    \mathcal{L} \supset Z^\prime_\mu \left(e \epsilon \, J^\mu_{\rm EM} +  \frac{g}{2 c_W}\frac{m_{Z^\prime}^2}{m_Z^2} \chi \, J^\mu_{\rm NC} + g_X \, J^\mu_{D}\right),
\end{equation}
where we assume $m_{Z^\prime}\simeq g_X v_\phi \ll m_Z$, and define $\epsilon\equiv c_W \chi$.

\section{Low Energy Anomalies}
Our aim is to show that the model can explain 
several low energy anomalies, while simultaneously generating the correct scale for light neutrino masses. Since mixing in the light neutrino sector can be generated by appropriate choices of the $M_D$ matrix, we work under the simplifying assumption of a single active neutrino generation, in our case $\nu_\mu$, denoting the lightest non-zero mass eigenstate by $\nu_3$. We require that $m_3 \simeq 0.05$ eV, compatible with the scale suggested by neutrino oscillation experiments~\cite{PDG2020}. 
For concreteness, we pick $n=3$ sterile and $d=1$ vector-like dark neutrinos, although only the three lightest heavy neutrino mass eigenstates $N_j$, $j=4,5,6$, will be important for the phenomenology we discuss. The heaviest states $N_7$ and $N_8$ have masses of several GeVs, and are mostly-sterile states. 

Our proposal is illustrated by two benchmark points (BPs), one exhibiting a left-right symmetry and one without. Their properties are shown in \reftab{tab:BPmain} but a detailed definition is left to \refapp{app:BPdef}. The left-right symmetry in the dark sector of BP-A ($\nu_{D_L}^c \leftrightarrow \nu_{D_R}$) is achieved by setting $Y_L = Y_R$, and explains the vanishing entries in \reftab{tab:BPmain}. This can be shown to be related to CP conservation.

Let us comment on the generic features of our two BPs. We fix $m_{Z^\prime}=1.25$~GeV and $\epsilon^2 =  4.6\times10^{-4}$ for the dark photon. The three lightest HNLs all have $\mathcal{O}(100)$~MeV masses, and decay via neutrino and kinetic mixing as $N_i \to N_{i-1} e^+e^-$. Specifically, $N_5$ will typically decay with $c\tau^0_5 \lesssim 3$~cm, leading to displaced $e^+e^-$ vertices, while $N_6$ will decay more promptly, $c\tau^0_6 \lesssim 3$~mm. In the case of $N_4$, it can only decay into SM particles, $N_4\to \nu e^+e^-$, making it much longer-lived, $c\tau^0_4 \lesssim 100$ km. In addition, $N_4$ is mostly sterile, which naturally leads to $\mathcal{B}(Z^\prime\to~N_4 N_4) \ll \mathcal{B}(Z^\prime \to N_{\{4,5,6\}} N_{\{5,6\}})$, and explains why $c\tau^0_6 < c\tau^0_5$. For concreteness, we fix $m_{\phi^\prime}=1$~GeV, forbidding fast $N_6 \to N_j \phi^\prime$ decays and respecting perturbativity limits on the dark scalar quartic coupling $\lambda_{\Phi}$.

\emph{$\Delta a_\mu$ and BaBar --} A discrepancy between the most precise $\Delta a_\mu$ measurement performed by the Muon $(g-2)$ collaboration~\cite{Bennett:2006fi} and theoretical calculations~\cite{Davier:2010nc,Davier:2017zfy,Blum:2018mom,Keshavarzi:2018mgv,Davier:2019can} stands at more than $3.7\sigma$~\footnote{Recent lattice calculations~\cite{Borsanyi:2020mff} predict values closer to the experiment. However, this has been pointed out to lead to inconsistencies with $e^+e^-\to$ hadrons data~\cite{1802519,Crivellin:2020zul}. For the latest consensus in this field, see \refref{Aoyama:2020ynm}}. In view of the efforts by the Muon $(g-2)$ collaboration to measure this quantity four times more precisely at FNAL~\cite{Grange:2015fou}, it is timely to reconsider the dark photon solution to the $\Delta a_\mu$ puzzle~\cite{Pospelov:2008zw}. Minimal dark photon models are excluded by collider and beam dump searches for $Z^\prime \to \ell^+\ell^-$~\cite{Lees:2014xha,Aaij:2019bvg,Bauer:2018onh,Fabbrichesi:2020wbt}. If a GeV dark photon decays invisibly, then it is subject to strong limits from monophoton searches at BaBar~\cite{Lees:2017lec}. This constrains $\epsilon^2 \lesssim 10^{-6}$ for $m_{Z^\prime}<3$~GeV by searching for a missing-mass resonance produced alongside initial-state radiation (ISR), $e^+e^-\to \gamma Z^\prime$. In models where the $Z^\prime$ decays semi-visibly inside the detector, $\mathcal{B}(Z^\prime\to$vis$+\slashed{E})\simeq1$, this limit can be relaxed. This was proposed in the context of inelastic DM models in Ref.~\cite{Mohlabeng:2019vrz}, and later criticized in a more conservative analysis~\cite{Duerr:2019dmv} (see also ~\cite{Essig:2009nc,Baumgart:2009tn, Batell:2009yf}).

In our model, however, the mechanism put forward in Ref.~\cite{Mohlabeng:2019vrz} is improved, as more visible energy is deposited in the detector. For the bound to be relaxed above the central value to explain the $\Delta a_\mu$ anomaly, the detection inefficiency for the $Z^\prime$ decay products in ISR events ought to be at most $0.22\%$. Note that in virtually all ISR monophoton events the produced $Z^\prime$ promptly decays into $N_{5}$ and/or $N_6$ states, which subsequently lead to one or more $e^+e^- + \slashed{E}$ vertices. Such additional particles are hard to miss in the barrel-like BaBar detector, which operates with a $1.5$ T magnetic field. In fact, after produced, all $N_6$ states decay already inside the drift chamber, while for BP-(A,B), we find that, for a typical $2.5$~GeV $N_5$ energy, $(79,92)\%$ of $N_5$ states decay before the electromagnetic calorimeter (ECAL), followed by $(11, 5.7)\%$ inside the ECAL, and $(8.0, 2.3)\%$ in the muon detection system. Fully invisible decays are rare and satisfy $\mathcal{B}(Z^\prime\to N_4 N_4) + \mathcal{B}(Z^\prime\to N_4 N_5)\times P_{N_5}^{\rm escape} \lesssim 2.2\times10^{-3}$ for the BPs. Visible decays are also negligible, $\mathcal{B}(Z^\prime\to \ell^+\ell) \lesssim \mathcal{O}(10^{-5})$. 

\begin{figure}
    \centering
    \includegraphics[width=0.46\textwidth]{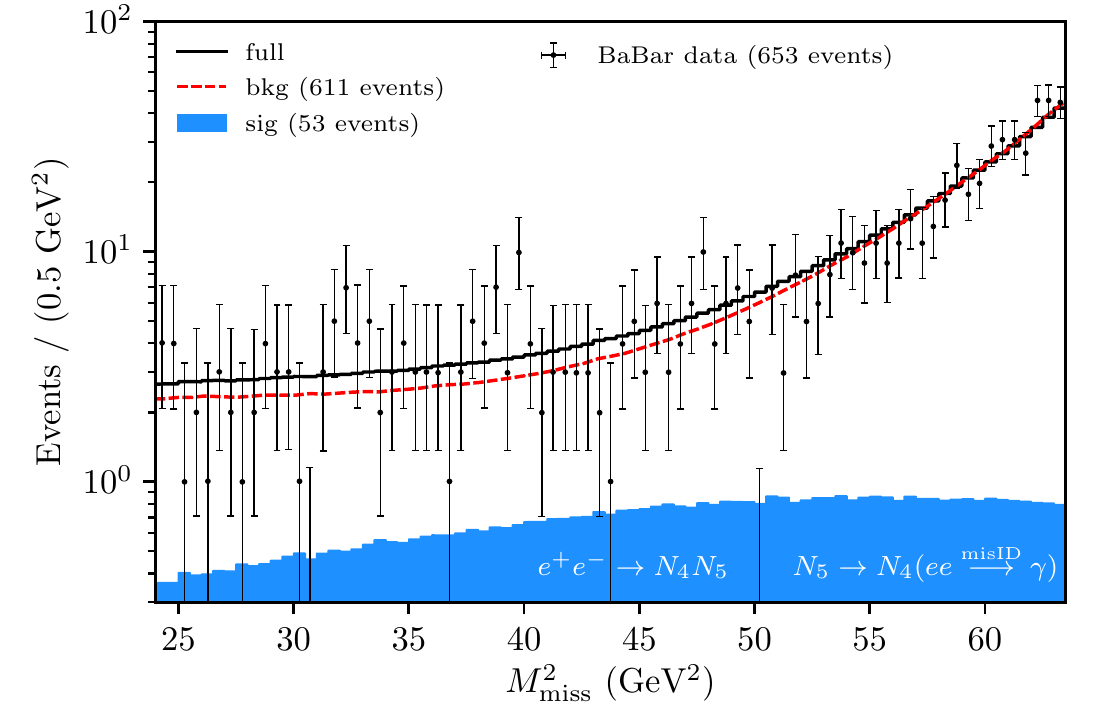}
    \caption{BaBar monophoton data at large $M_{\rm miss}^2=s-2 E_{\gamma}^* \sqrt{s}$. The background prediction quoted by the collaboration (red) is added to the best fit prediction in our BP-B (blue) in the solid black line. Event numbers are for entire HighM region ($24$~GeV$^2<M_{\rm miss}^2 < 64$~GeV$^2$).}
    \label{fig:BaBar_data}
\end{figure}

\emph{Pseudo-monophotons at BaBar --} The dominant production of dark particles in $e^+e^-$ colliders is s-channel pair production of HNLs due to the large values of $\alpha_D \epsilon^2$. In particular, the process
\begin{equation}
    e^+e^- \to Z^{\prime*} ( \text{or }\Upsilon(nS)) \to N_4 (N_5\to N_4 e^+e^-),
\end{equation}
could fake monophoton signatures when the $N_5$ decays inside the BaBar ECAL. These events could contribute to the large missing mass ($M_{\rm miss}^2 \equiv s - 2\sqrt{s} E_\gamma^{CM}$) monophoton sample at BaBar, where we point out that a mild excess is observed in the $24$~GeV$^2 < M_{\rm miss}^2<50$~GeV$^2$ region.

For an integrated luminosity at BaBar of 15.9~fb$^{-1}$ in $\sqrt{s}=10.02$~GeV and $22.3$~fb$^{-1}$ in $\sqrt{s}=10.36$~GeV, BP-(A,B) predict a total number of single pseudo-monophoton events of
\begin{align}\label{eq:HNLrate}
(3.6,\,9.6) \times 10^4 \times \mathcal{P}_{N_5}^{\gamma}  \times \epsilon_{\rm B},
\end{align}
where $\epsilon_{\rm B}$ is the final detection and selection efficiency of the monophoton analysis at BaBar, not including the probability $\mathcal{P}_{N_5}^{\gamma}$ that the $N_5$ states decay inside the ECAL \emph{and} get reconstructed as a photon. For the ISR analysis, $\epsilon_{\rm ISR}\simeq 0.2-3.5\%$, depending on $M_{\rm miss}^2$. In our pseudo-photon case, however, it is impossible to estimate $\epsilon_{\rm B}$ without a dedicated detector simulation and the machine learning algorithm utilized by BaBar. Nevertheless, we fit our model prediction to data, which will give an estimate of the value of $\mathcal{P}_{N_5}^{\gamma}\epsilon_{\rm B}$ required to explain the excess in the model. Since backgrounds are much larger than our signal above $M_{\rm miss}^2 > 50$~GeV$^2$, our fit uses only the data in $24 $~GeV$^2 < M_{\rm miss}^2 < 50$~GeV$^2$, where a total of $189$ events are observed on top of a prediction of $157$ background events. Floating $\mathcal{P}_{N_5}^{\gamma}\epsilon_{\rm B}$ for BP-B, we minimize a binned Poisson likelihood, assigning a $1\,(5)\%$ normalization uncertainty on the background model. We find a  $2.5\sigma\,(2.2\sigma)$ preference for 53 signal events. Our best-fit point in BP-B is shown in \reffig{fig:BaBar_data}, where events were selected if $\theta_{ee}<10^\circ$, and the boost and azimuthal angle cuts were implemented as in Ref.~\cite{Lees:2017lec}. This corresponds to a total number of $93$ pseudo-monophoton events, before any selection cuts. Finally, since both BPs predict very similar shapes, we can make use of Eq.~\ref{eq:HNLrate} to find $\mathcal{P}_{N_5}^{\gamma}\epsilon_{\rm B}\simeq (0.26,0.10)\%$. A dedicated analysis at BaBar would be able to determine if such numbers are experimentally justified. We also note that our pseudo-monophoton rate is compatible with constraints on $\mathcal{B}(\Upsilon(1S)\to \gamma + \slashed{E}) < 5.6 \times 10^{-6}$ at $90\%$~C.L. at BaBar~\cite{delAmoSanchez:2010ac}, provided the $e^+e^-\to\gamma$ mis-ID rate is less than $(100,77)\%$ for BP-(A,B).

\begin{figure*}[t]
    \centering
    \includegraphics[width=0.49\textwidth]{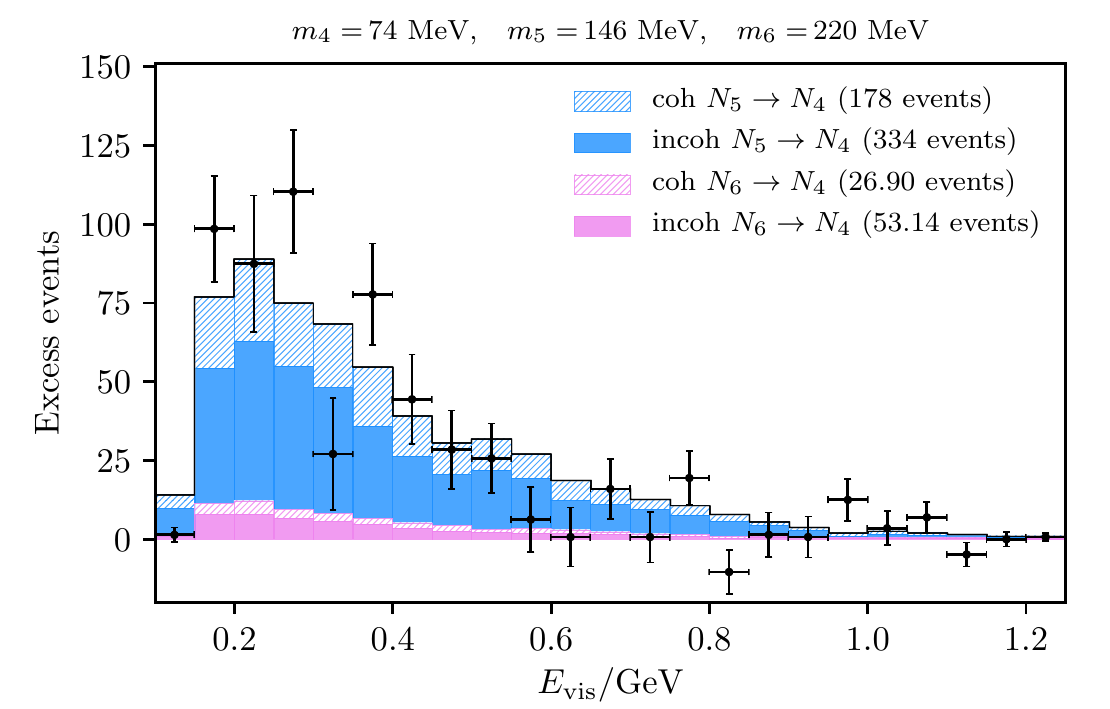}
    \includegraphics[width=0.49\textwidth]{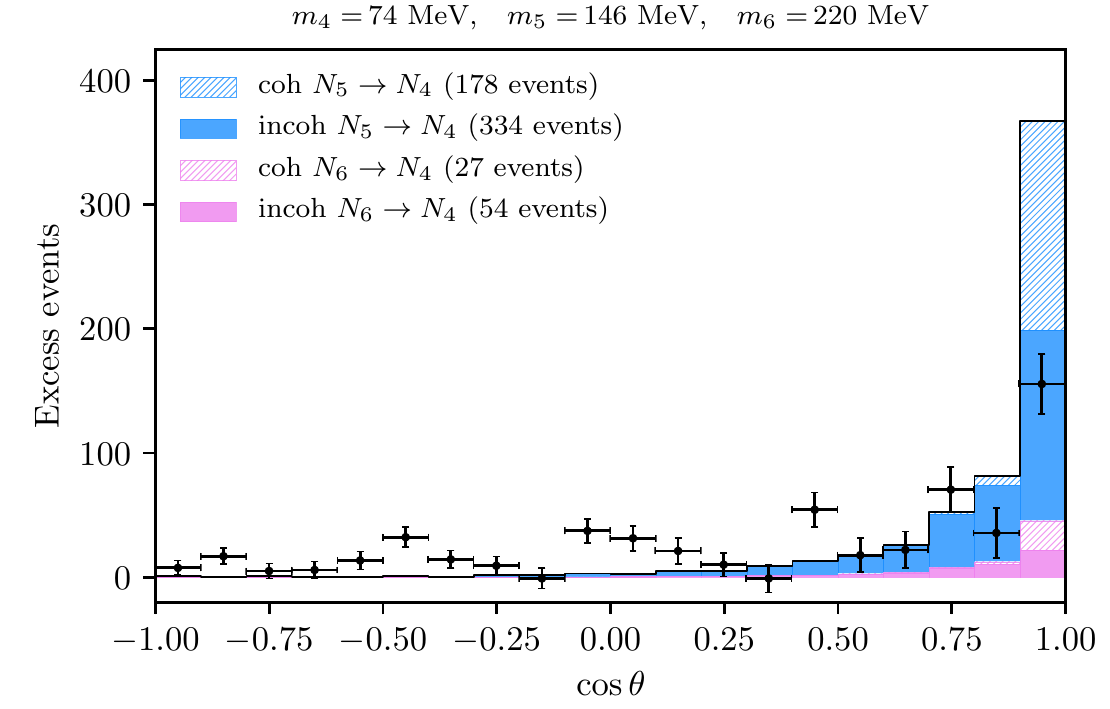}
    \caption{MiniBooNE low energy excess and our model prediction in BP-B for $\nu_\mu$ upscattering into $N_5\to N_4 e^+e^-$ (blue) and $N_6\to N_4 e^+e^-$ (pink) in BP-B. The incoherent (filled) and coherent (hashed) scattering contributions are shown separately.\label{fig:MBevents}}
\end{figure*}

\emph{MiniBooNE excess --} MiniBooNE is a mineral oil Cherenkov detector in a predominantly $\nu_{\mu}$ beam with $\langle E_\nu \rangle \simeq 800$ MeV. Recent results with improved background analysis and larger statistics~\cite{Aguilar-Arevalo:2020nvw} report an excess of $560.6\pm119.6$ ($77.4\pm28.5$) $e$-like events in $\nu$ ($\overline{\nu}$) mode. Initially designed to search for short-baseline oscillations reported by the LSND experiment~\cite{Aguilar:2001ty}, MiniBooNE reports a much more significant $4.8\sigma$ excess. 
The large tensions with global datasets in oscillation models~\cite{Adamson:2020jvo,Aartsen:2020fwb,Aartsen:2020iky} (see also~\cite{Dentler:2018sju,Diaz:2019fwt,Boser:2019rta}) prompts new scenarios to explain the excess.

We propose that the MiniBooNE excess arises from the decay products of HNLs produced in $\nu_\mu$ upscattering inside the detector,
\begin{equation}
    \nu_\mu \,+\, \mathcal{H} \,\to\, (N_{6,5} \to N_4 \,+\,e^+\,+\,e^-) \,+\, \mathcal{H},
\end{equation}
where $\mathcal{H} = \{C, p^+ \}$ is the hadronic target. The $e^+e^-$ pairs with small angular aperture or large energy asymmetry mimic a single EM shower in the Cherenkov detector. This is similar to the upscattering explanation proposed in Ref.~\cite{Ballett:2018ynz}, but successfully achieves fast HNL decay without infringing upon any bounds~\footnote{In Refs.~\cite{Gninenko:2009yf,Gninenko:2009ks,Gninenko:2010pr,Gninenko:2011hb}, a similar idea was proposed in the context of a transition magnetic moment, which closely resembles the light dark photon models later studied in Refs.~\cite{Bertuzzo:2018itn,Ballett:2018ynz,Arguelles:2018mtc}. Such scenarios predict exclusively forward signatures, $\cos{\theta}> 0.95$. Other models with scalars decaying to $e^+e^-$ have been discussed in Refs.~\cite{Datta:2020auq,Dutta:2020scq,Abdallah:2020biq,Liu:2020ser}.}.

A prediction of our signal on top of MiniBooNE neutrino data is shown in \reffig{fig:MBevents} for our BP-B. In our single generation approximation, the upscattering cross section is proportional to $|V_{3 j}|^2 \alpha_D (e \epsilon)^2$, where $|V_{3j}|^2\equiv |U_{D_L 3}^*U_{D_L j} - U_{D_R 3}^*U_{D_R j}|^2$ is the mixing factor in the $\nu_3 N_j Z^\prime$ interaction and takes $\mathcal{O}(10^{-7})$ values.
The scattering is predominantly electromagnetic via $Z^\prime$ exchange, and due to the large values of $\alpha_D \epsilon^2$, no interference with the SM $Z$ is observed. This, together with the purely vectorial couplings of the $Z^\prime$, explains why the signal prefers to be forward with respect to charge-current quasi-elastic scattering. We note that scattering on protons is dominant, and that the angular spectrum predictions can improve when nuclear effects and higher $Q^2$ scattering regimes are included. The produced $e^+e^-$ that contributes to the excess has a small invariant mass, with $m_{ee}<m_{5,6}-m_4$. If $m_{ee}$ is too large, it contributes to the NC $\pi^0$ dataset, where an excess is also observed~\cite{AguilarArevalo:2010xt}. We estimate the overall detection and signal selection efficiency for our BPs to be $\simeq 5\%$. Although many upscatterings lead to $N_6\to (N_5\to N_4 e^+e^-) e^+e^-$, we do not include these double vertex events as a large fraction of them would be excluded by the MiniBooNE cuts.

\emph{Old accelerator anomalies --} 
Many accelerator experiments in the 80s and 90s searched for $\nu_\mu \rightarrow \nu_e$ transitions at short-baselines, with some of them observing significant excesses. While a neutrino oscillation interpretation of these results is excluded, they can be explained within our model, where the energy dependence and signal characteristics differ from those of oscillation. The largest deviation was observed by the PS-191 experiment at CERN using a $E_{\nu}^{\rm peak} \sim 600$~MeV $\nu_\mu$ beam and the fine-grained ECAL component of their detector. They observed an excess of $23\pm 8$ $e$-like events on a background of $12\pm 3$ events, amounting to a $3\sigma$ significance~\cite{Bernardi:1986hs,Bernardi:1987ek}. All excess events contained a scattering vertex, followed by an electromagnetic shower $<16$~mm away. A follow-up experiment, E-816~\cite{Astier:1989vc}, was designed to test the PS-191 anomaly at the Brookhaven National Laboratory (BNL) with a wide-band beam of mean energy $\langle E_\nu\rangle\simeq 1.5$~GeV. E-816 also reported an excess of $e$-like events with a small vertex-shower separation of $<8.8$~mm, although at a lower significance of $\gtrsim 2\sigma$ due to larger systematic errors~\cite{Astier:1989vc}. In our model, these excesses can be explained by $\nu_\mu$ upscattering to $N_6$, which decays very fast to overlapping or energy-asymmetric $e^+e^-$ pairs, fitting the exponential drop of events as a function of vertex-shower separation. PS-191 and E-816 observed a larger excess than MiniBooNE, which could be explained by the larger $N_6$ upscattering rate (BP-B) or solely due to different signal reconstruction (BP-A).

Other experiments with $E_\nu^{\rm peak} \gtrsim 1$~GeV reported no excess, namely E-734~\cite{Ahrens:1984gp} and E-776~\cite{Blumenfeld:1989qm,Borodovsky:1992pn}. The stringent cuts against $\pi^0$ backgrounds would veto most of our $e^+e^-$ pairs and weaken the constraint. Another set of bounds come from high energy experiments, such as NOMAD~\cite{Astier:2003gs}, with $\langle  E_\nu \rangle \simeq 24$~GeV, and CCFR~\cite{Romosan:1996nh} and NuTeV~\cite{Avvakumov:2002jj}, both with $\langle  E_\nu \rangle \simeq 140$~GeV. Their bounds, although very strong under the oscillation hypothesis, are much weaker for our model due to the $\log{E_\nu}$ growth of the $Z^\prime$ mediated neutrino-nucleus cross-sections in comparison to the linear $E_\nu$ growth in the SM. Finally, we note an unexplained excess of positron events observed at NOMAD~\cite{Astier:2003gs} in a sideband sample of events containing showers far from the scattering vertex or that had failed kinematic cuts. 
Such positrons are predicted in our model as coming from asymmetric $e^+e^-$pairs in the late decays of our HNLs.

We also note an intriguing excess reported by CCFR in the search for HNLs produced in scattering~\cite{Mishra:1987xh,deBarbaro:1990sq,Budd:1992vt,deBarbaro:1992hb}. The experiment saw evidence for double-vertex events with $9$ NC/NC events over an estimated overlay background of $3\pm 0.2 \, (\mbox{stat.}) \pm 0.4 \, (\mbox{syst.})$. A double-vertex event was defined as one in which there were ``two distinct and separate shower regions'', and NC/NC refers to two neutral vertices, as opposed to NC/CC events, wherein a second vertex contained a muon candidate. No excess was observed in the NC/CC, which disfavored standard interpretations with HNLs that have large branching ratios to muons. In our model, only NC/NC events appear, mainly from upscattering into $N_6$, which immediately decays into $N_5 e^+ e^-$, with the subsequent $N_5 \rightarrow N_4 e^+ e^-$ decays typically happening after a few meters at CCFR energies. This leads to good agreement with the $4$ to $14$~m vertex-shower separation observed, given the typical $N_5$ energies of $50$~GeV. A naive scaling of the cross sections shows that the normalization is compatible with the rate at MiniBooNE and PS-191.

\section{Discussion and Conclusions} 
Let us remark that our BPs satisfy all existing experimental constraints, including decay-in-flight bounds from PS-191~\cite{Bernardi:1985ny,Bernardi:1987ek}. Searches for peaks in the muon spectrum in $\pi^+/K^+\to \mu^+ N_j$~\cite{Artamonov:2014urb, CortinaGil:2017mqf} are also satisfied due to strong vetoes against visible energy in the detector, as discussed in Ref.~\cite{Ballett:2019pyw}. Intriguingly, the latest results from $K^+ \to e^+ N_j$ searches at NA62~\cite{NA62:2020mcv} indicate an excess at $m_N=346$~MeV, with $|U_{ej}|^2\simeq 1.5 \times 10^{-9}$ at 2.2$\sigma$ (3.6$\sigma$)  global (local) significance. Our model can accommodate this hint by identifying $N_5$ with the required HNL and switching on the mixing with the electron neutrinos. To take into account the visible decays of our HNL, the required $|U_{ej}|^2$ is enhanced by a factor $\sim2$ for $\sim5$~ns lifetimes, as quoted by the experiment. 
For our BPs, we also expect to see an excesses in the muon sector, depending on the $K^+\to\mu^+ N_{j}$ efficiency at NA62.

There is some freedom in the choice of the HNL parameters while keeping the same key phenomenological features, e.g. HNL decay length and $Z^\prime$ branching ratios. For the dark photon parameters, the situation is more constrained. For instance, lower values of $m_{Z^\prime}$, such as $1$~GeV with $\epsilon^2 = 3\times10^{-4}$ are possible, and decrease the required $\alpha_D |V_{3j}|^2$ couplings to explain MiniBooNE, PS-191, and BaBar by a factor of $(1.25)^4\sim 2.5$. Going much below $m_{Z^\prime}=1$~GeV leads to more forward angular distribution at MiniBooNE and introduces tension with neutrino-electron scattering constraints~\cite{Arguelles:2018mtc}. A survey of existing bounds and additional BPs are provided in Appendices \ref{app:bounds} and \ref{app:BPdef}.

We also want to highlight the left-right symmetry in BP-A, as in that case the lightest HNL $\nu_4$ has vanishing interactions with the $Z^\prime$, except for the $|V_{45}|^2$ vertex. Incidentally, $N_4$ could lie at the keV scale, and may be a candidate for non-thermal dark matter~\cite{Dodelson:1993je}.

%
Direct searches for our MiniBooNE explanation can be performed at the Short-Baseline Neutrino program at FNAL~\cite{Antonello:2015lea,Machado:2019oxb}, which comprises three Liquid Argon detectors: SBND, $\mu$BooNE, and ICARUS. Specifically, for BP-(A,B) we predict that $\mu$BooNE~\cite{Acciarri:2016smi} would see a total number of $\sim760$ neutrino upscattering events into $N_5$ and $(0,2800)$ events into $N_6$, before any efficiencies and for a total $N_{\rm POT} = 13.6\times10^{20}$. While the former would contain a single $e^+e^-$ pair, the latter events would constitute double vertex events with $\gtrsim 10$~cm separation. Around $60\%$ of the total number of events are due to coherent scattering, and leave no visible proton tracks. Dedicated studies of the $e^+e^-$ invariant mass, as well as searches for the double-vertex events would help discriminate our hypothesis from other dilepton MiniBooNE explanations. Other near detectors to experiments like MINER$\nu$A, NO$\nu$A, and T2K could also shed light on the model. In particular, the incoherent piece of our prediction may be constrained by photon-like showers in $\nu_e$ CC quasi-elastic measurements and the coherent piece by photon-like showers at neutrino-electron scattering measurements. Searches for $e^+e^-$ pairs in decay-in-flight at the ND280, the off-axis near detector of T2K, can also constrain HNLs produced in coherent neutrino upscattering inside the detector~\cite{Brdar:2020tle}.

Other direct searches can be performed at the NA62 kaon facility~\cite{NA62:2017rwk}. The decays of $75$~GeV/c kaons to $K^+\to \ell_\alpha^+ N_i$ followed by $N_j\to N_k e^+e^-$ would constitute a background-free signature, similar to the one proposed in Ref.~\cite{Ballett:2019pyw}. The new physics events would appear as a displaced $e^+e^-$ vertex with peaked kinematics, where $(p_K - p _\ell)^2 = m_{j}^2$, $(p_K-p_\ell - p_{ee})^2=m_k^2$, and $p_{ee}^2 = (p_{e^-}+p_{e^+})^2\leq (m_j - m_k)^2$. The production rate is controlled by $|U_{\mu j}|^2$, where for BP-(A,B) we predict a total $K^+\to\mu^+ (N_6\to N_5 e^+e^-)$ event rate of $(1970,2980)$ for $N_K = 2.14\times 10^{11}$ fiducial kaon decays and an overall $4\%$ acceptance~\cite{CortinaGil:2019dnd,CortinaGil:2019nuo}. 

The dark photon can be searched for in the ISR events at BaBar, Belle-II~\cite{Kou:2018nap,Duerr:2019dmv}, and BESIII~\cite{Zhang:2019wnz} by relaxing the vetoes on additional $e^+e^-$ pairs in the detector. The large value of $\epsilon^2$ required for the $\Delta a_\mu$ explanation yields several hundred events at BaBar. Direct $N_j N_k$ pair production, as well as Higgstrahlung $e^+e^-\to \phi^\prime Z^\prime$, would also appear as multiple displaced $e^+e^-$ vertices at $B$-factories, and in the fixed-target experiments NA64~\cite{NA64:2019imj,Gninenko:2020hbd} and LDMX~\cite{Akesson:2018vlm}, providing a background-free signature for semi-visible dark photons. 

In summary, this \emph{letter} provides an explanation to some of the most prominent low energy anomalies, including the MiniBooNE excess and the $\Delta a_\mu$ anomaly. 
The phenomenological signatures we presented are achieved in a renormalizable model which extends the SM by an anomaly-free $U(1)^\prime$ gauge symmetry and a dark neutrino sector. The model is able to reproduce the correct scale for the light neutrinos, albeit with some level of fine tuning.
Phenomenologically, our scenario only requires a semi-visible GeV-scale dark photon that couples to $\mathcal{O}(100\text{ MeV})$ HNLs. We show that the dark photon not only evades sensitive searches for missing mass resonances at BaBar, but can actually explain a mild but continuous excess seen in the data thanks to the pseudo-monophotons from $N_5\to N_4 e^+e^-$ decays. Due to the large kinetic mixing required by $\Delta a_\mu$, such events naturally arise from s-channel $e^+e^-$ collisions producing HNLs. We point out that $e$-like events from upscattering are better able to explain past anomalies reported by PS-191 and E-816, compared to those from excluded oscillation hypotheses. Also curious is the prediction of $\mathcal{O}(2~\text{cm})$ lifetime for $N_5$, as it leads to double vertex events at neutrino experiments and is compatible with a significant excess reported by CCFR.
The novel interplay between portal couplings and exotic decay signatures in our model offer striking signatures at current and upcoming experiments. Observations of displaced vertices at kaon and neutrino experiments, as well as the decays of a semi-visible dark photon, would provide confirmation of our model.

\begin{acknowledgments}
It is a pleasure to acknowledge discussions with Carlos Arg\"uelles, Martin Bauer, Evgueni Goudzovski, and Mike Shaevitz. We thank Maxim Pospelov for discussions and for pointing out additional constraints in an earlier version of this draft.
This project has received partial funding from the European Union’s Horizon 2020 research and innovation programme under the Marie Sklodowska-Curie grant agreement No. 690575 (RISE InvisiblesPlus) and No. 674896 (ITN Elusives) and the European Research Council under ERC Grant NuMass (FP7-IDEAS-ERC ERC-CG 617143). The research at the Perimeter Institute is supported in part by the Government of Canada through NSERC and by the Province of Ontario through Ministry of Economic Development, Job Creation and Trade, MEDT.
AA is funded by the UKRI Science, Technology and Facilities Council (STFC).
\\
\\
\end{acknowledgments}

\appendix
\section{Details on Benchmark Points}\label{app:BPdef}

In the main text we have focused only on the phenomenological aspects of our model, giving two BPs that can resolve the low energy anomalies. In this appendix, we offer more details on the model side, giving the vertex factors for each relevant interaction that can be used to compute physical observables. The BPs in the main text were given in terms of a model with a single generation of active neutrinos, $n=3$ sterile states and $d=1$ dark vector-like fermions. 
We also present two additional BPs to illustrate the ranges of the HNL masses compatible with the phenomenology discussed. In particular, BP-C indicates the smallest scale of $m_6$ and $m_5$ which lead to sufficiently fast $N_6$ and $N_5$ decays. With BP-D, we illustrate the features of heavier masses. 

Following Eq.~\ref{eq:mass_matrix} in the main text, the full mass matrix is given as,
\begin{align}
\frac{1}{2}\overline{\makevec{\nu}_f^c} \left( 
    \begin{matrix} 0 & M_{D1} &  M_{D2} &  M_{D3} & 0 & 0 \\ 
                   M_{D1} & M_1 & 0 & 0 & \Lambda_{L1} & \Lambda_{R1}\\
                   M_{D2} & 0 & M_2 & 0  & \Lambda_{L2} & \Lambda_{R2}\\
                   M_{D3} & 0 & 0 & M_3  & \Lambda_{L3} & \Lambda_{R3}\\
                   0 & \Lambda_{L1} & \Lambda_{L2} & \Lambda_{L3} & 0 & M_X \\
                   0 & \Lambda_{R1} & \Lambda_{R2} & \Lambda_{R3} & M_X & 0 \end{matrix}
    \right)\makevec{\nu}_f \,,
\end{align}
where now $\makevec{\nu}_f \equiv \left( \begin{matrix} \makevec{\nu}_\alpha^c & \makevec{\nu}_{N_1}^c & \makevec{\nu}_{N_2}^c & \makevec{\nu}_{N_3}^c & \makevec{\nu}_{D_L}^c& \makevec{\nu}_{D_R} \end{matrix} \right)^T$. The values for the mass matrix parameters used for our BPs are given in \reftab{tab:theory_params}. 

In the mass basis, HNLs mixing with the different flavors is given in \reftab{tab:mixings}. To clarify the nature of our neutrino couplings to the neutral bosons, we write the explicit vertices in the neutrino mass basis using the flavor gauge boson basis. To leading order in $\chi$ and taking light dark photons: $Z_\mu = Z^0_\mu + s_W \chi X_\mu$ and $Z^\prime_\mu = X_\mu - s_W \chi Z^0_\mu$. The interactions are given by
\begin{align}
    \mathcal{L}_{\rm int} &\supset      \frac{g}{2c_W} Z_\mu^0 \overline{\makevec{\nu}_m} \gamma^\mu \frac{(C P_L - C^\dagger P_R)}{2} \makevec{\nu}_m 
    \\\nonumber
    &\quad + g_X X_\mu  \overline{\makevec{\nu}_m} \gamma^\mu \frac{(V P_L - V^\dagger P_R)}{2} \makevec{\nu}_m 
    \\\nonumber
    &\quad + h \,\overline{\makevec{\nu}_m} \frac{(H P_L + H^\dagger P_R)}{2\sqrt{2}} \makevec{\nu}_m
    \\
    &\quad+ \phi \,\overline{\makevec{\nu}_m} \frac{(S P_L + S^\dagger P_R)}{2\sqrt{2}} \makevec{\nu}_m, \nonumber
\end{align}
where $\makevec{\nu}_{m}$ is the mass eigenvector and $P_{L,R} = (1\mp\gamma^5)/2$. The vertex factors are defined as
\begin{align}\label{eq:vertexfactors}
    C &= U_{\alpha}^\dagger U_{\alpha},
    \\
    V &= U_{D_L}^\dagger U_{D_L} - U_{D_R}^\dagger U_{D_R},\nonumber
    \\
    H &= U_N^T Y U_{\alpha}+ U_{\alpha}^T Y^T  U_N, \nonumber
    \\
    S &= U_N^T (Y_L U_{D_L} + Y_R U_{D_R})+ (Y_L U_{D_L} + Y_R U_{D_R})^T U_N. \nonumber
\end{align}
We show the relevant vertex factors for dark bosons in our BPs in \reftab{tab:vertexfactors}. For all BPs, we take $m_{Z^\prime}=1.25$ GeV, $m_{\phi^\prime}=1$ GeV and $\epsilon^2 = 4.6 \times 10^{-4}$. The mixing $\sin{\theta}^2$ is assumed to be negligible for our BPs. The HNLs with masses above $m_{Z^\prime}$, namely $N_7$ and $N_8$, are mostly in the sterile direction, with $|V_{jk}|^2 \ll 1$, and $|U_{N_2 j}|^2, |U_{N_3 j}|^2 \sim \mathcal{O}(1)$ for $j=7,8$.

The phenomenology of BP-C is similar to BP-B, with $c\tau^0_5\simeq 2.2$~cm and $c\tau^0_6\simeq 1.2$~mm. Notably, it represents the smallest scale of HNL masses with lifetimes that are compatible with the old accelerator anomalies, although it requires a slightly larger $\alpha_D$. This point also allows for the lightest scalar $\phi^\prime$.
On the other hand, BP-D features considerably larger masses with the largest $N_5$ lifetime, $c\tau^0_5\simeq 4$~cm, and smallest $N_4$ lifetime of any point, $c\tau^0_4\simeq 4$~km. Displaced vertices would be slightly enhanced here, although the heavier masses result in slightly worse distributions at MiniBooNE, more peaked at lower energies.

As mentioned in the main text, our model is also compatible with hints of a mild excess at NA62 and we illustrate this with BP-D. We identify the $346$~MeV HNL as $N_5$, and turn on mixing with the electron neutrinos. Taking the Yukawa couplings in the electron sector as $Y_e \simeq~ 0.11~Y_\mu$, or $M^e_{D_i} \simeq 0.11~M^\mu_{D_i}$ for $i=(1,2,3)$, we obtain the mixings
\begin{align}
    |U_{e 4}|^2 &\simeq 0,
    \nonumber\\
    |U_{e 5}|^2 &\simeq 3.00 \times 10^{-9}, \nonumber\\
    |U_{e 6}|^2 &\simeq 4.59 \times 10^{-9}.
\end{align}
It is important to note that the bounds from NA62 on both $|U_{e 5}|^2$ and $|U_{e 6}|^2$ are weakened due to the fast decays of $N_5$ and $N_6$. For $N_5$ with lifetimes $\sim5$~ns, the experiment expects a weakening of the bound by a factor $\sim2$~\cite{NA62:2020mcv} implying an effective $|U_{e 5}|_{\rm NA62}^2\simeq 1.5\times10^{-9}$.

The electron mixing requires two active light neutrinos, $\hat{\nu}_2$ and $\hat{\nu}_3$. With our chosen Yukawas, $\hat{\nu}_2$ is massless and mostly in the $\nu_e$ direction, with $\hat{\nu}_3$ mostly in the $\nu_\mu$ direction. As we do not consider the full $8\times8$ mass matrix with three active light neutrinos, we do not attempt to reproduce the structure of the PMNS matrix, but note that this can be achieved with appropriate choices of the Yukawa couplings in the active sub-block. The scattering cross-section at MiniBooNE is now proportional to $\sum_{i=2,3}|U_{\mu i}V_{ij}|^2\alpha_D (e \epsilon)^2 \simeq |U_{\mu 3}|^2|V_{3j}|^2 \alpha_D (e \epsilon)^2$, since the $|V_{2j}|$ mixings are negligible for massless $\hat{\nu}_2$.

\renewcommand{\arraystretch}{1.3}
\begin{table*}[t]
    \centering
    \scalebox{0.9}{
    \begin{tabular}{|c|cccc|c|c|c c c|c c c|c c c c c c| c c c|}
    \hline
    \multirow{2}{*}{BP} & \multirow{2}{*}{MB} & \multirow{2}{*}{$\Delta a_\mu$} & \multirow{2}{*}{BB}& \multirow{2}{*}{Acc} & \multirow{2}{*}{\,$\alpha_D$\,}  & $m_3$ & $m_4$& $m_5$ & $m_6$ & $|V_{43}|^2$ & $|V_{53}|^2$ & $|V_{63}|^2$ & \multicolumn{6}{c|}{$\mathcal{B}(Z^\prime \to N_j N_k)/\%$} & \multicolumn{3}{c|}{$c\tau^0$/cm}
    \\
     & & & & &\, &\, /eV \,& \multicolumn{3}{c|}{/MeV} & \multicolumn{3}{c|}{$/10^{-8}$} & 44 & 45 & 46 & 55 & 56 & 66 & $N_4$ & $N_5$ & $N_6$ \\
    \hline \hline
    {A} & \checkmark
    & \checkmark & \checkmark
    & (\checkmark) &  $0.39$ & $0.05$ & $35$ & $120$ & $185$ & $0$ & $22.2$ & $0$ & $0$ & $5.4$ & $0$ & $0$ & $95$ & $0$ & $1.6 \times 10^{13}$ & $3.0$ & $0.26$ \\
     {B} & \checkmark & \checkmark & \checkmark & \checkmark &  $0.32$ & $0.05$ & $74$ & $146$ & $220$ & $13.6$ & $26.5$ & $123$ & $0.15$ & $11$ & $0.48$ & $1.6$ & $86$ & $0.59$ & $1.1 \times 10^7$ & $2.2$  & $0.14$\\
    {C} & \checkmark & \checkmark & \checkmark & \checkmark & $0.76$ & $0.05$ & $62$ & $110$ & $180$ & $13.7$ & $11.2$ & $33.2$ & $0.00014$ & $30$ & $0.019$ & $0.23$ & $70$ & $0.15$ & $1.1 \times 10^7$ & $2.2$ & $0.12$ \\
   {D} & \checkmark & \checkmark & \checkmark & (\checkmark) & $0.11$ & $0.05$ & $275$ & $346$ & $435$ & $1.44$ & $75.3$ & $17.1$ & $0.021$ & $13$ & $0.060$ & $0.13$ & $87$ & $0.023$ & $4.0 \times 10^5$ & $3.9$  & $0.13$\\
    \hline

    \end{tabular}}
    \caption{Illustrative benchmark points (BP-C and BP-D). For ease of comparison, we report also the values for BP-A and BP-B. For all points, $m_{Z^\prime} = 1.25$ GeV. Here, the $V_{ij} \equiv U_{D_L i}^*U_{D_L j} - U_{D_R i}^*U_{D_R j}$ are the mixing factors in $Z^\prime N_i \nu_j$ vertices, and $\alpha_D = g_X^2 /4\pi$. Note that $Z^\prime \to \nu_3 \nu_3$ is negligible for the mixings considered here. We refer to the MiniBooNe excess as MB, the BaBar excess as BB, 
    and the accelerator experiments as Acc. The zeroes in BP-A are protected by a left-right symmetry ($\Lambda_L = \Lambda_R$).}
    \label{tab:BPmainAPP}
\end{table*}

\begin{table}[t]
\centering
\begin{tabular}{|c c|c c c c|}
    \hline
    \multicolumn{6}{|c|}{Table of Theory Parameters}\\
    \hline
     & & {A} & {B} & {C} & {D}\\
    \hline \hline
    $m_{D 1}$ & \multirow{3}{*}{$/10^{6}$ eV}
    & $0.00950$ & $-0.0347$ & $0.0336$ & $0.129$\\
    $m_{D 2}$ & & $0.278$ & $1.98$ & $-0.635$ & $6.72$\\
    $m_{D 3}$ & & $0.190$ & $-3.89$ & $-1.03$ & $-11.4$ \\
    \hline \hline
    $M_1$ & \multirow{3}{*}{$/10^{9}$ eV}
    & $-0.0429$ & $-0.0900$ & $-0.0963$ & $0.206$\\
    $M_2$ & & $1.10$ & $6.00$ & $5.07$ & $6.00$\\
    $M_3$ & & $-1.10$ & $-18.0$ & $-10.1$ & $-18.0$\\
    \hline \hline
    $\Lambda_{L 1}$ & \multirow{3}{*}{$/10^{7}$ eV} & $-2.39$ & $3.75$ & $3.51$ & $15.0$\\
    $\Lambda_{L 2}$ & & $19.0$ & $24.0$ & $25.5$ & $33.0$\\
    $\Lambda_{L 3}$ & & $0.00$ & $0.00$ & $12.7$ & $0.00$\\
    \hline \hline 
    $\Lambda_{R 1}$ & \multirow{3}{*}{$/10^{7}$ eV} & $-2.39$ & $-2.81$ & $-4.04$ & $-14.9$\\
    $\Lambda_{R 2}$ & & $19.0$ & $54.0$ & $44.1$ & $-12.9$\\
    $\Lambda_{R 3}$ & & $0.00$ & $0.00$ & $-38.1$ & $0.00$\\
    \hline \hline
    $M_X$ & $/10^{8}$ eV & $-1.21$ & $1.96$ & $1.56$ & $3.50$\\
    \hline
    \end{tabular}
    \caption{Theory parameters for $1+3+1$ model.\label{tab:theory_params}}
\end{table}

\begin{table}[h]
\centering
\begin{tabular}{|c c|c c c c|}
    \hline
    \multicolumn{2}{|c|}{Neutrino mixing} & {A} & {B} & {C} &{D} \\
    \hline \hline
    $|U_{\mu 4}|^2$ & \multirow{3}{*}{$/10^{-8}$} & $45.5$ & $0.00361$ & $0.000256$ & $0$ \\
    $|U_{\mu 5}|^2$ & & $0$ & $157$ & $51.1$ & $22.7$ \\
    $|U_{\mu 6}|^2$ & & $8.28$ & $14.0$ & $12.8$ & $69.5$ \\
    \hline \hline
    $|U_{N_1 4}|^2$ & \multirow{3}{*}{$/10^{-2}$} & $94.9$ & $88.8$ & $71.3$ & $90.2$\\
    $|U_{N_1 5}|^2$ & & $0$ & $0.162$ & $0.0139$ & $0.0303$\\
    $|U_{N_1 6}|^2$ & & $5.14$ & $11.1$ & $28.7$ & $9.75$\\
    \hline \hline
    $|U_{N_2 4}|^2$ & \multirow{3}{*}{$/10^{-4}$} & $27.3$ & $2.79$ & $1.91$ & $3.30$\\
    $|U_{N_2 5}|^2$ & & $0$ & $83.4$ & $96.2$ & $5.86$\\
    $|U_{N_2 6}|^2$ & & $398$ & $12.8$ & $5.45$ & $23.06$\\
    \hline \hline
    $|U_{N_3 4}|^2$ &  \multirow{3}{*}{$/10^{-4}$} & $0$ &  $0$ & $3.65$ & $0$\\
    $|U_{N_3 5}|^2$ & & $0$ & $0$ & $2.75$ & $0$\\
    $|U_{N_3 6}|^2$ & & $0$ & $0$ & $9.51$ & $0$\\
    \hline \hline
    $|U_{D_L 4}|^2$ & \multirow{3}{*}{$/10^{-1}$} & $0.244$ & $0.371$ & $1.43$ & $0.554$\\
    $|U_{D_L 5}|^2$ & & $5.00$ & $5.57$ & $5.19$ & $4.83$\\
    $|U_{D_L 6}|^2$ & & $4.54$ & $4.04$ & $3.36$ & $4.59$\\
    \hline \hline
    $|U_{D_R 4}|^2$ & \multirow{3}{*}{$/10^{-1}$} & $0.244$ & $0.749$ & $1.44$ & $0.421$\\
    $|U_{D_R 5}|^2$ & & $5.00$ & $4.33$ & $4.71$ & $5.16$\\
    $|U_{D_R 6}|^2$ & & $4.54$ & $4.84$ & $3.76$ & $4.41$\\
    \hline 
    \end{tabular}
    \caption{Neutrino mixing parameters for our BP-A, B, and C. Note that $U_{D_L} = U_{D_R}$ for BP-A due to $\Lambda_L = \Lambda_R$.\label{tab:mixings}}
\end{table}

\begin{table}[h]
\centering
\begin{tabular}{|c c|c c c c|}
    \hline 
     \multicolumn{2}{|c|}{$Z^\prime$ vertex} & {A} & {B} & {C} &{D}\\
     \hline \hline
    $|V_{43}|^2$ & \multirow{3}{*}{$/10^{-8}$}
    & $0$ & $13.6$ & $13.7$ & $1.44$\\
    $|V_{53}|^2$ & & $22.2$ & $26.5$ & $11.2$ & $75.3$\\
    $|V_{63}|^2$ & & $0$ & $123$ & $33.2$ & $17.1$\\       
    \hline \hline
    $|V_{44}|^2$ & \multirow{6}{*}{$/10^{-3}$}
    & $0$ & $1.43$ & $0.00134$ & $0.176$\\
    $|V_{45}|^2$ & & $48.7$ & $105$ & $284$ & $96.7$\\
    $|V_{46}|^2$ & & $0$ & $4.60$ & $0.184$ & $0.530$\\
    $|V_{55}|^2$ & & $0$ & $15.2$ & $2.23$ & $1.10$\\
    $|V_{56}|^2$ & & $909$ & $869$ & $702$ & $899$\\
    $|V_{66}|^2$ & & $0$ & $6.31$ & $1.59$ & $0.300$\\
\hline \hline
    \multicolumn{2}{|c|}{$\phi^\prime$ vertex} & {A} & {B} & {C} &{D}\\
\hline\hline
    $|S_{33}|^2$ & $/10^{-14}$ & $0.205$ & $7.87$ & $2.17$ & $1.30$\\
\hline\hline
    $|S_{43}|^2$ & \multirow{3}{*}{$/10^{-8}$} & $0.0926$ & $0.675$ & $0.273$ & $5.33$\\
    $|S_{53}|^2$ & & $0$ & $1.31$ & $0.509$ & $0.00670$\\
    $|S_{63}|^2$ & & $0.163$ & $0.0598$ & $17.4$ & $1.62$\\       
    \hline \hline
    $|S_{44}|^2$ & \multirow{6}{*}{$/10^{-2}$}
    &  $0.0203$ & $0.0783$ & $0.841$ & $0.708$\\
    $|S_{45}|^2$ & & $0.0305$ & $0.394$ & $0.0853$ & $0.00108$\\
    $|S_{46}|^2$ & & $0.181$ & $0.418$ & $11.3$ & $4.71$\\
    $|S_{55}|^2$ & & $0$ & $1.20$ & $1.58$ & $0.000959$\\
    $|S_{56}|^2$ & & $0.444$ & $0.744$ & $50.0$ & $0.107$\\
    $|S_{66}|^2$ & & $0.668$ & $0.258$ & $10.1$ & $1.47$\\
    \hline
    \end{tabular}
    \caption{The vertex factors entering in $Z^\prime \nu_i N_j$ ($|V_{ij}|^2$) and $Z^\prime N_j N_k$ ($|V_{j k}|^2$) interactions, as well as in $\phi^\prime \nu_i N_j$ ($|S_{ij}|^2$) and $\phi^\prime N_j N_k$ ($|S_{j k}|^2$) interactions, as defined in \refeq{eq:vertexfactors}. \label{tab:vertexfactors}}
\end{table}
    
\section{Survey of Existing Constraints} 
\label{app:bounds}
\paragraph{Electroweak precision observables} 
An assessment of the impact of kinetic mixing on electroweak precision observables (EWPO) requires a global fit to collider and low energy data. This was performed in Ref.~\cite{Curtin:2014cca}, where a model independent bound on $\epsilon$ was derived. For $m_{Z^\prime} \ll M_Z$, the authors find $\epsilon^2_{\rm EWPO} < 7.3\times 10^{-4}$ at $95\%$ C.L, just above our value of $\epsilon^2 = 4.6\times 10^{-4}$. As a sanity check against more recent data, we also directly compute the oblique parameters $S$, $T$, and $U$~\cite{Peskin:1991sw} to leading order in $\epsilon = c_W \chi$ and $\mu\equiv g_X v_\phi/M_Z^{\rm SM}$, neglecting the impact of running in the dark couplings and corrections from dark fermion loops. For all our BPs, these are~\cite{Holdom:1990xp,Babu:1997st,Frandsen:2011cg}
\begin{align}
    S &\simeq 4 s_W^2 \epsilon^2 (1+\mu^2)/\alpha = 0.042,
    \\
    T &\simeq - s_W^2 \chi^2 \mu^2/\alpha = -3.3\times10^{-6},
    \\
    U &\simeq  4 s_W^4 \epsilon^2/\alpha = 0.013.
\end{align}
Clearly, this is compatible with the current bounds of $T<0.22$ and $S<0.14$ at 95\% C.L.~\cite{PDG2020}. The constraints on $S$ can be much stronger when fixing $T=0$ and $U=0$, which would be mostly driven by the $1$ to $2\sigma$ discrepancies observed between direct $M_W$ measurements and the global best fit point. We plan to return to this issue in future communication~\cite{future}.

\paragraph{Deep-inelastic scattering constraints} 
Recently, Ref.~\cite{Kribs:2020vyk} appeared setting new model-independent constraints on dark photons using $e p^+$ scattering data from HERA~\cite{Abramowicz:2015mha}. At $95\%$ C.L., the authors find that $\epsilon^2 \lesssim 2.9\times 10^{-4}$, in tension with our BPs. As the authors discuss, inclusion of other datasets weakens the bound, which signals a mild tension between HERA and other experiments. Finally, we note that a naive rescaling of the constraints on contact interactions performed by the ZEUS collaboration~\cite{Abramowicz:2019uti}, where the probability distribution functions were included in the fit, leads to bounds that are weaker by a factor of $\sim 2$ than the ones quoted by Ref.~\cite{Kribs:2020vyk}. While HERA is certainly sensitive to our model at some level, we believe that it is not excluding it at the $95\%$ C.L.
Nevertheless, a trivial modification to our setup to accommodate such bound is to lower $m_{Z^\prime}=1$~GeV.

\paragraph{$Z\to$ invisible} Dark fermions can be produced in the decays of SM-like $Z$ bosons via its couplings to the dark current. This is induced by kinetic mixing, and to leading order in $\chi$ it is
\begin{equation}
\mathcal{L} \supset Z_\mu
 g_X s_W \chi J_X^\mu.
\end{equation}
This coupling is relevant in our model since $g_X \epsilon$ is not so small. A constraint can be derived from LEP measurements of the $Z$ boson decay width~\cite{ALEPH:2005ab} and constrains $\Gamma_{Z\to{\rm inv}} < 2$ MeV. The largest new physics decay mode is $Z \to N_j N_k$, for $j,k>3$, which even without requiring the HNLs to be invisible, yields
\begin{align}
    \Gamma_{Z\to N_j N_k} &\simeq \frac{ |V_{jk}|^2 G_F m_Z^3}{12 \sqrt{2}\pi}  \left(\frac{2 g_X s_W \epsilon}{g}\right)^2  
    \\\nonumber
    &\simeq 0.17 \text{ MeV }  \left(\frac{ \alpha_D|V_{jk}|^2 \epsilon^2}{4.6\times 10^{-4}}\right),
\end{align}
safely below the current constraints even for the largest $\alpha_D$ couplings, as it can be shown that $\sum_{j,k}^8 |V_{jk}|^2 = 2$.

Another relevant process is $Z\to Z^\prime \phi^\prime$. Neglecting the final state masses, we find
\begin{align}
    \Gamma_{Z\to Z^\prime \phi^\prime} &= \frac{\pi\alpha_D t_W^2\epsilon^2 M_Z}{12} \simeq 70\text{ keV} \left(\frac{ \alpha_D\epsilon^2}{4.6\times 10^{-4}}\right),
\end{align}
also satisfying the constraints independently of the fate of $\phi^\prime$ and $Z^\prime$ in the detector.
\paragraph{$h\to$ invisible} Searches for Higgs decays to invisible have been performed by CMS~\cite{Sirunyan:2018owy} and ATLAS~\cite{Aaboud:2019rtt}. Latest preliminary results by ATLAS require that $\mathcal{B}(h\to$invisible$) < 0.13$ at 95\% C.L.~\cite{ATLAS:2020cjb}, which for the SM value $\Gamma_h^{\rm SM} \simeq 4.07$ MeV, implies $\Gamma_{h\to \text{inv}} < 0.52$ MeV. This constrains the Yukawas and scalar parameters of the theory. 

Firstly, we consider $h \to N_i N_j$ neglecting scalar mixing. Saturating the bound, we find
\begin{align}
\Gamma_{h \to N_j N_k} & \simeq  \frac{s_\theta^2 |H_{jk}|^2m_h}{4\pi} 
= 0.52\text{ MeV } \left( \frac{|H_{jk}|^2 }{5.3\times10^{-5}}\right)
\end{align}
which does not lead to strong constraints on our model given that the largest Yukawa we have is for BP-D, where it is a few $10^{-5}$.

The parameters in the scalar potential are also subject to constraints. We consider $h^\prime \to \phi^\prime \phi^\prime$ induced by the scalar portal coupling $\lambda_{\Phi H}$. A direct computation neglecting final state masses leads to,
\begin{align}\label{eq:phiphi_bound}
    \Gamma_{h\to \phi^\prime \phi^\prime} &\simeq \frac{\lambda_{\Phi H}^2 v_h^2}{32\pi m_h}
    =0.52 \text{ MeV} \times \left(\frac{\lambda_{\Phi H}}{1.04\times 10^{-2}} \right)^2,
\end{align}
which can be interpreted as a strong constraint on our scalar mixing angle $\theta \simeq  (\lambda_{\Phi H}/2\lambda_H)\times(v_\Phi/v_H) < 8.1 \times 10^{-5}(v_{\Phi}/500 {\text{ MeV}})$. This value is currently an order of magnitude below current sensitivity of $K_L\to \pi^0 \phi^\prime$ searches at KOTO or $K^+\to \pi^+ \phi^\prime$ searches at NA62.

Decay to a pair of HNLs can also proceed via the dark Yukawas if scalar mixing is present. Neglecting the final state masses, we find
\begin{align}
    \Gamma_{h \to N_j N_k} &\simeq  \frac{s_\theta^2 |S_{jk}|^2m_h}{4\pi} 
    = 370\text{ eV } |S_{jk}|^2 \left( \frac{s_\theta}{8.1\times10^{-5}}\right)^2,
\end{align}
where $s_\theta^2 \equiv \sin^2 \theta$ is chosen according to \refeq{eq:phiphi_bound} for $v_{\phi}=500$~MeV. This clearly satisfies the bound as it can be shown that $\sum_{j,k}^8 |S_{j,k}|^2 = 4 (|Y_L|^2 + |Y_R|^2)$. Similarly, the decay $h\to Z^\prime Z^\prime$ is possible and may appear invisible some fraction of the time. Nevertheless, the tree-level rate is 
\begin{align}
    \Gamma_{h\to Z^\prime Z^\prime} &\simeq \frac{\alpha_D s_\theta^2 m_h}{4 \pi} = 86 \text{ eV } \alpha_D \left( \frac{s_\theta}{8.1\times10^{-5}}\right)^2,
\end{align}
and loop-corrections from fermion loops are also negligible.

\emph{$K\to\pi \phi^\prime$} The KOTO experiment at J-PARC~\cite{Ahn:2018mvc} has set stringent constraints on $\mathcal{B} (K_L\to\pi^0 \slashed{E})$. Initial hints of a signal in the latest unblinding~\cite{kotoKAON19} have later been revisited due to unexpected charged kaon backgrounds and signal mis-identification~\cite{kotoFPCP20,kotoICHEP2020}. The hinted values led to branching ratios much larger than the SM prediction~\cite{Buras:2015qea}, and prompted several new physics studies~\cite{Kitahara:2019lws,Egana-Ugrinovic:2019wzj,Dev:2019hho,Liu:2020qgx,Cline:2020mdt}. 
In light of the new backgrounds and given stringent constraints on the scalar mixing found above, we refrain from trying to explain these events. We note that KOTO, as well as NA62, are only sensitive to singlet scalar emission in $K \to \pi \phi^\prime$ decays for mixings of order $s^2_{\theta} \sim 10^{-7}$, which for our small $v_{\Phi}$ are already excluded due to $h\to \phi^\prime \phi^\prime$ decays. A more exotic scalar sector could be invoked to explain this excess, where a new invisible real scalar $S$ could be hidden under background at NA62 if $m_S\sim m_\pi$~\cite{Fuyuto:2014cya}, and could lead to signatures at KOTO without violating the Grossman-Nir constraints~\cite{Grossman:1997sk}.

\paragraph{Meson $\to$ invisible} 

We consider the decays of vector meson states due to the vector nature of the dark photon couplings. The best current bounds are at the level of $\mathcal{B}(J/\psi \to$ inv)$ < 7.2 \times 10^{-4}$ at BES~\cite{Ablikim:2007ek}, and 
$\mathcal{B}(\Upsilon(1S) \to$ inv)$ < 3.0 \times 10^{-4}$ at BaBar~\cite{Aubert:2009ae}. The branching ratios into HNLs, $\mathcal{B}(V\to N_j N_k)$, may be still be slightly above such values, provided a sufficient number of the produced HNLs decay semi-visibily. In general, the branching ratio for the quarkonium states (V) used throughout our article is
\begin{align}\label{eq:Vbrs}
\mathcal{B}&(V \to N_j N_k) = \frac{\alpha |V_{jk}|^2 (g_X \epsilon Q)^2 \tau_{V}}{3} \frac{M_{V}^3 f_{V}^2}{(M_{V}^2 - m_{Z^\prime}^2)^2}  \nonumber
\\
&={|V_{jk}|^2 \alpha_D} \times \left\{ \begin{matrix} 
5.5\times 10^{-3} \text{   for } V=J/\psi\phantom{1S}
\\  
1.7 \times 10^{-3} \text{   for } V=\Upsilon(1 S) 
\\
1.3  \times 10^{-3} \text{   for } V =\Upsilon(2 S) 
\\  
1.5 \times 10^{-3} \text{   for } V=\Upsilon(3 S) 
\\  
0.86 \times 10^{-6} \text{   for } V=\Upsilon(4 S)  \end{matrix}\right.,
\end{align}
where we neglected the final state masses, and took $m_{Z^\prime} = 1.25$~GeV and $\epsilon^2= 4.6\times 10^{-4}$. The decay constants, $f_{\Upsilon(nS)} = 498, 430, 336$~MeV for $n=2,3,4$, were extracted from existing $V\to e^+e^-$ measurements in Ref.~\cite{Tanabashi:2018oca}. For the fully invisible states $N_4 N_4$ (as well as for light neutrinos) the mixing factor $V_{jk}$ is sufficiently small to avoid the constraints. Production of $N_4 N_5$ is the next largest contribution and it still satisfies the most stringent limits from BES, since in all BPs $|V_{45}|^2 \times P_{N_5}^{\rm escape} < 10\%$, where $P_{N_5}^{\rm escape}$ is the probability for $N_5$ to escape detection.

\paragraph{Pseudo-monophotons} As discussed in the main text, s-channel $e^+e^-$ collisions at BaBar can lead to pseudo-monophoton events. It is not possible to extract a constraint from this without a dedicated detector simulation, as it relies on experimental details such as the efficiency to reconstruct our $e^+e^-$ as a photon, and on the specifics of the machine learning algorithm.
In the main text, however, we proceeded to understand if it is at all feasible to explain a mild excess observed in the monophoton data. By finding a best-fit value for the normalization of events that are ``photon-like", we have asked whether such rate is possible within our model and whether the efficiencies it requires are reasonable. Here, photon-like refers to events where a $N_4 N_5$ pair is produced in the interaction point, followed by a $N_5\to N_4 e^+e^-$ decay inside the ECAL. When plotting our prediction, we required that the angle of separation between the electrons be less than $\theta_{ee}<10^\circ$, and select events within the angular acceptance of the ECAL detector. For our total pair-production rate, we include HNLs produced in $Z^\prime$ mediated s-channel $e^+e^-$ collisions, where the $e^+e^- \to (Z^\prime)^* \to N_i N_j$ cross section was found to be
\begin{align}
\frac{\dd \sigma_{e^+e^- \to N_i N_j}}{\dd \cos{\theta_{\rm CM}}} &\simeq |V_{i j}|^2 \alpha \alpha_D \epsilon^2   \frac{\pi s}{(s-m_{Z^\prime}^2)^2} \frac{ (1+ \cos^2{\theta_{\rm CM}})}{2},
\end{align}
neglecting final state masses and where $\theta_{\rm CM}$ is the center of mass angle between $N_i$ and the collision axis. We also include a contribution from  $e^+e^-\to \Upsilon(nS)$ ($n=2,3,4$), followed by decay into HNLs. We use $\sigma_{ee\to \Upsilon(2S)}(\sqrt{s} = 10.02\text{ GeV}) \simeq 7$~nb  and $\sigma_{ee\to \Upsilon(3S)}(\sqrt{s} = 10.36\text{ GeV}) \simeq 4$~nb as well as \refeq{eq:Vbrs}.

\paragraph{Higgstrahlung} Another source of HNL production at $e^+e^-$ colliders is dark higgstrahlung. Due to the large dark coupling, the process $e^+e^- \to \phi^\prime Z^\prime$ is important~\cite{Batell:2009yf}, with a differential cross section
\begin{equation}
    \frac{\dd \sigma_{e^+e^- \to \phi^\prime Z^\prime}}{\dd \cos{\theta_{\rm CM}}} \simeq \pi \alpha\,\alpha_D \epsilon^2  \frac{\sin^2{\theta_{\rm CM}} (s-m_{Z^\prime})^2 + 8 m_{Z^\prime}^2 s}{4 s^2 (s - m_{Z^\prime}^2)},
\end{equation}
where $\theta_{\rm CM}$ is the center of mass angle between the $Z^\prime$ and the collision axis. This process is therefore comparable to direct HNL production. Remaining agnostic about the decay products of $\phi^\prime$, but requiring the decay $Z^\prime \to N_j N_k$, the ratio between direct and higgstrahlung HNL production in our BPs is
\begin{align}
&\frac{4 |V_{jk}|^2 s^3}{(s-m_{Z^\prime}^2)((s-m_{Z^\prime}^2)^2 + 12 s\, m_{Z^\prime}^2)} \times \frac{1}{\mathcal{B}(Z^\prime \to N_j N_k)},
\\\nonumber
&\quad \simeq 3.5\times\frac{|V_{jk}|}{\mathcal{B}(Z^\prime \to N_j N_k)}.
\end{align}
For production of $N_4N_5$ pairs in ours BPs A and B, this constitutes a ratio of just above $3$. Given that $\phi^\prime$ decays promplty and visibly, especially into $N_6$ states, we do not include this contribution in our monophoton discussion, but emphasize that this offers yet more visible signatures at $e^+e^-$ colliders.

\paragraph{$\Upsilon(1S) \to$ invisible + $\gamma$} Vector meson decays to photon plus missing energy are direct probes of our pseudo-monophoton events. The full process is $\Upsilon(2S)\to \pi^+\pi^- (\Upsilon(1S)\to\gamma+\slashed{E})$, where the $\pi^+\pi^-$ kinematics can be used to identify the $\Upsilon(1S)$ state. The current limits are quoted in terms of the BR into an invisible pseudoscalar, $\Upsilon(1S) \to \gamma +A^0$, and into a pair of invisible fermions, $\Upsilon(1S) \to \gamma \chi\overline{\chi}$. Most relevant to us are the three-body decay limits taken at the smallest $\chi$ masses ($m_{\chi}\to 0$), where BaBar~\cite{delAmoSanchez:2010ac} constrains 
\begin{align}
\mathcal{B}&(\Upsilon(1S) \to \gamma \chi \chi)<5.6\times10^{-6},
\end{align}
which was improved by Belle~\cite{Seong:2018gut} to
\begin{align}
\mathcal{B}&(\Upsilon(1S) \to \gamma \chi \chi)<3.5\times10^{-6},
\end{align}
all at the 90\% C.L. More recently BESIII~\cite{Ablikim:2020qtn} has set the strongest limits on the two-body process 
\begin{align}
\mathcal{B}(J/\psi \to \gamma A^0)<6.9\times10^{-7},
\end{align}
but since the missing mass in this process is fixed, the constraint does not apply to us.

When implementing these bounds on our model, we used the BRs in \refeq{eq:Vbrs}. For comparing the $\Upsilon(1S)\to\gamma+\slashed{E}$ 
constraints to the BaBar pseudo-monophoton rate, only the BaBar limit is taken into account, as $\mathcal{P}_{N_5}^{\gamma}$ is a detector-dependent quantity, and, under the simplifying assumption that it is constant in energy and angle, it is the same for the two processes.

Since the efficiencies for our pseudo-monophoton events are different than in the s-channel production mode, we use the limits above to obtain an upper-bound on the detector-dependent quantity $\mathcal{P}_{N_5}^{\gamma}$. Neglecting the sub-dominant $N_5 N_5$ contribution, BaBar sets a limit of $\mathcal{B}(\Upsilon(1S) \to N_4 N_5) \times \mathcal{P}_{N_5}^{\gamma} < 5.6 \times 10^{-6}$ at 90\% C.L, which implies $\mathcal{P}_{N_5}^{\gamma} < (18,10,1.5,30)\%$ for BP-(A,B,C,D). Since the probability for $N_5$ to decay inside the ECAL is known to be $(14,13,14,10)\%$ for BP-(A,B,C,D) ($E_{N_5} \simeq 5$~GeV for s-channel production), we use the limit on $\mathcal{P}_{N_5}^{\gamma}$ to find the largest allowed $e^+e^-\to\gamma$ mis-ID rate for our explanation of the BaBar excess not to be excluded. Under the approximation that it is independent of the kinematics, this rate is bounded from above by $(100,77,12,100)\%$. Clearly this allows to explain the monophoton excess while remaining consistent with the $\Upsilon(1S)\to\gamma\slashed{E}$.

\paragraph{NA64 searches} The fixed-target NA64~\cite{NA64:2019imj,Gninenko:2020hbd} experiment also sets stringent limits on invisible dark photons. The $100$~GeV electrons can produce dark photons via bremsstrahlung in interactions with the dense beam dump material, $e \text{W}\to e \text{W} Z^\prime$, with $\text{W}$ a Tungsten nucleus. The bound from $2.84\times 10^{20}$ electrons on target is shown in Ref.~\cite{NA64:2019imj} for invisible dark photons with masses as large as $m_{Z^\prime}\sim0.94$~GeV. For the latter dark photon mass, it constrains $\epsilon^2 \lesssim 10^{-4}$ at 90\% C.L. As discussed in Ref.~\cite{Gninenko:2020hbd}, the semi-visible decays of the dark photon can weaken the bound. For our BPs ($m_{Z^\prime} = 1.25$~GeV), we do not have access to the exact value of the invisible-$Z^\prime$ constraint, but under a conservative assumption of linear scaling with $m_{Z^\prime}$, we find that NA64 does not constrain our BPs provided $\sim45\%$ of $Z^\prime$ particles produced are vetoed due to the subsequent semi-visible decays of $N_{5,6}$. For the HNLs produced in the decay of the highest energy dark photons ($E_{N_{5,6}}\sim 50$~GeV), we have a typical decay length $c\tau$ of $\mathcal{O}(10\text{ m})$ and $\mathcal{O}(1\text{ m})$ for $N_5$ and $N_6$, respectively. Note that the dark photons are produced inside the beam dump and decay promptly. Under these conservative assumptions, we find that most HNLs decay before or within the instrumented ECAL of NA64, assumed to be a total of $\sim10$~m. The presence of one or more vertices of $e^+e^-$ pairs would be vetoed from the invisible-$Z^\prime$ search due to visible showers in the ECAL, as well as in the additional veto detectors.

\paragraph{Beam dump and decay-in-flight searches}\label{app:HNLdecay} HNLs heavier than $N_4$ in our model are unconstrained by decay-in-flight searches due to their short lifetimes. On the other hand, $N_4$ is longer-lived and faces strong constraints from HNL searches at PS-191~\cite{Bernardi:1985ny,Bernardi:1987ek}. If $N_4$ has new interactions, such as in BP-B, it decays faster than in the minimal HNL models, and the constraints from decays in-flight are modified (see, \eg, the discussion in Ref.~\cite{Ballett:2016opr}). While $N_4$ is produced in $\pi,K\to \mu N_4$ decays, which are controlled by $|U_{\mu4}|$, its subsequent $N_4\to\nu e^+e^-$ decays in BP-B proceed mainly through $Z^\prime$ exchange, which is controlled by $|V_{43}|$. In that case, we require
\begin{equation}
    |U_{\mu 4}|^2|V_{43}|^2 < {|U^*_{\mu 4}U_{e 4}|_{\text{PS}}^2}\left(\frac{\sqrt{2} G_F m^2_{Z^\prime}}{e\epsilon g_X}\right)^2 \times \text{F}\,,
\end{equation}
where $|U^*_{\mu 4}U_{e 4}|_{\text{PS}}$ is the bound quoted by the PS-191 experiment. The factor $\text{F}= 3.17$ converts the bound on Dirac to Majorana HNLs, and takes into account that PS-191 assumed only charged-current decays in their analysis.

We note that there are additional production channels for $N_4$ than in the minimal HNL models. On top of the standard meson decays $\pi,K \to \ell N$, HNLs can also be produced via kinetic mixing in $\rho, \omega \to N_i N_j$ and $\pi^0,\eta, \eta^\prime \to \gamma N_i N_j$, where the vector meson decays dominate. These channels have been explored in the context of two-component fermionic dark sectors in Refs.~\cite{Gninenko:2012eq,Ilten:2018crw,Tsai:2019mtm,Darme:2020ral}. In this context, limits on new dark sector fermions that decay to $e^+e^-+\slashed{E}$ have been set using CHARM~\cite{Bergsma:1985qz} and NuCal~\cite{Blumlein:1990ay} data. With the effective field theory approach of Ref.~\cite{Darme:2020ral}, we see that for $c\tau_{N_i} \simeq 10$ cm or smaller such constraints are safely avoided due to the short lifetimes. For the longer-lived $N_4$, however, a simplified re-scaling of the constraints from Ref.~\cite{Darme:2020ral}, where $g^2/\Lambda^4 \to |V_{45}| |U_{\mu4}| G_F (e\epsilon g_X/2m_{Z^\prime}^2)$, and $g^2/\Lambda^4 \to |V_{45}| |V_{43}| (e\epsilon g_X/2m_{Z^\prime}^2)^2$ for BPs B and C, shows that the constraints are satisfied. A re-analysis of the PS-191 constraints including these new production mechanisms for $N_4$, both from vector meson decays as well as from the secondary decays of $N_5, N_6$ could set stronger constraints in our parameter space, but is beyond the scope of this work.

We would like to highlight an event found in PS-191 and shown in Fig.~9 of Ref.~\cite{Chauveau:1985iz}: it has two tracks in the initial decay detector which subsequently shower in the ECAL. While it is unlikely to be due to photons, as they would not be recorded in the flash tubes, it could be due to two electrons coming from an $N_4$ decay. We do not elaborate further on this intriguing event.

\paragraph{Peak searches -- } Searches for a missing mass in $\pi/K\to \mu N_j$ decays set stringent limits on $|U_{\mu 4}|^2$. In our model, the $N_5$ and $N_6$ states can be produced, but would lead to visible signatures inside the two most relevant experiments, namely E949~\cite{Artamonov:2014urb} and NA62~\cite{CortinaGil:2017mqf,NA62:2020mcv}. As pointed out in Ref.~\cite{Ballett:2019pyw}, the small probability to miss additional energy deposition in these experiments together with the stringent vetoes against $K\to\mu\nu\gamma^{(*)}$ backgrounds would, in fact, veto most of our events. For E949, the detection inefficiency is estimated to be larger than $0.5\%$, the typical photon inefficiency, and so the constraints would be weakened by factors of $\gtrsim 200$. A similar argument can be made about NA62, where the $e^+e^-$ signatures would have to be missed by several different detector components. In this way, our BPs are not excluded by peak searches, in particular BP-B where we rely on a relaxation by a factor of $\sim 10$ on the E949 limit on $|U_{\mu6}|^2$. Note that the mass of both $N_4$ and $N_5$ is always below the mass interval constrained by both E949 and NA62. This is important as $N_5$ has a larger probability to escape these detectors due to its $\mathcal{O}(10~\text{m})$ decay lengths in the laboratory frame.

\paragraph{Lepton number violating searches}
Due to electron mixing, we predict large rates for neutrinoless double-beta decay in BP-D. In addition to the light neutrinos, heavy $N_j$ states also contribute and may dominate. Their contribution contains large uncertainties as the momentum dependence of the nuclear matrix element is important at the $\mathcal{O}(100)$~MeV mass scale. Nevertheless, a naive rescaling of the effective mass
\begin{equation}
m_{\beta \beta} \simeq \left|\sum_{i=1}^{8} \frac{m_i U_{ei}^2}{1+  m_i^2/\langle p^2\rangle}\right|^2,
\end{equation}
leads to $m_{\beta\beta} \approx 130$~meV for $\langle p^2\rangle = (100 \text{ MeV})^2$ when all matrix elements $U_{ei}$ are real. This is to be compared with the current experimental sensitivity of $m_{\beta\beta} < 61 - 165$~meV at KamLAND-Zen~\cite{KamLAND-Zen:2016pfg}.
We interpret this as a suggestion that, unless strong cancellations due to the Majorana phases are at play, we predict observable rates of neutrinoless double beta within current experimental reach. Loop contributions~\cite{LopezPavon:2012zg} and a full three active flavor treatment of the mixing matrix are also important and should be studied in more detail.

Lepton number violating kaon decays of the type $K^+\to\mu^+ (N_j\to\mu^+\pi^-)$ are important for generic heavy Majorana neutrinos~\cite{Atre:2009rg,Abada:2017jjx,CortinaGil:2019dnd}. This signature proceeds via charged-current branching ratios of $N_j$, and it is much suppressed in our model whenever $j>4$, where $\mathcal{B}(N_j\to \mu^+\pi^-)\sim 10^{-8} - 10^{-7}$. For $N_4$, such decays can in principle have large branching ratios, but the long-lifetimes of $N_4$ renders the experimental searches insensitive.

\section{Old Accelerators Experiments -- Additional Details}\label{app:accels}

We now provide additional details regarding the accelerator experiments. 

\paragraph{PS-191}\label{app:PS191norm}
The PS-191 detector was made of $5\times5$~mm$^2$ flash tube chambers interleaved with $3$~mm thick iron plates giving the detector a very fine granularity, $3$~mm of iron or $\sim17\%$ of a radiation length. This was used to distinguish photons, whose showers started further from the vertex due to conversions, from electrons, which showered immediately. As shown in Fig.~3 of Ref.~\cite{Bernardi:1986hs}, most single-shower events started within the first chamber, which corresponded to $\sim 16$~mm. The analysis was restricted to events above $400$~MeV, to avoid $\pi^0$ backgrounds. The initial $e$-like shower sample contained a total 57 events, which, after cuts on energy and distance between vertex and shower start, left a pion background of $7\pm3$ events.

In our model, the most frequent upscattering events at PS-191 would produce a $N_6$, which immediately decays into $N_5 e^+e^-$. The electromagnetic (EM) shower created by this decay is then close to the upscattering vertex, and would explain the sharp drop in number of events as a function of the shower-vertex distance. The issue of normalization between the number of events required at PS-191 versus those observed at MiniBooNE, which is between a factor $5$ to $7$ larger, can be explained by the fact that not all $N_6$ events in MiniBooNE count as signal. This is due to the additional decay of $N_5$, which yields a total of four charged leptons that are very rarely mis-reconstructed as a single EM shower. Therefore, by increasing $|V_{63}|^2$ in comparison to $|V_{53}|^2$, one can increase the ratio of signal events between PS-191 and MiniBooNE. This happens in BP-B, where $|V_{63}|^2 \gtrsim 4.6 |V_{53}|^2$, although it is forbidden in BP-A due to the left-right symmetry. It should be noted that the coherent cross section in Iron is larger than in Carbon, and that the most energy-asymmetric $e^+e^-$ pairs may be reconstructed as a one track plus one shower events. 

\paragraph{E-816}\label{app:E816}

The E-816 experiment used the same fine-grained ECAL as PS-191. The number of events was quoted as a function of the scattering vertex and the start of the shower, allowing for $e/\gamma$ differentiation. This was measured in units of 0.25 radiation lengths ($\sim17.6$~mm). Any photons converting before this would fake electrons, although the exponential nature of the conversion makes such events unlikely.
The experiment searched for $\nu_e$ excesses in the one track one shower ($1$T$1$S) sample. To reduce the $\pi^0$ background, showers with $E\lesssim300$~MeV were cut from the analysis. According to their simulations, such cuts eliminated $\sim70\%$ of $\pi^0$s while only removing $\sim10\%$ of $\nu_e$s. After cuts, the $\pi^0$ background dropped to $\sim1.6\%$ of the $\nu_\mu$ interactions and was of the order of the $\nu_e$ contamination in the beam. The electron excess was then given by the subtraction of the $1$T$1$S events due to pions and those due to intrinsic $\nu_e$ background from the remaining $1$T$1$S events. They found an excess of $43\pm17.8\text{ (stat.)}\pm9\text{ (sys.)}$ and quoted a significance of $2.4\pm0.5~\sigma$.

Similar to PS-191, E-816 would also count upscattering events into $N_6$ as signal when the $e^+e^-$s are overlapping or highly energy-asymmetric. The ratio of $\nu_e$-like events to $\nu_\mu$-like events, $R = (\nu_e+\overline{\nu_e})/(\nu_\mu+\overline{\nu_\mu})$, observed at E-816, $R_{\rm observed}/R_{\rm expected} =1.6\pm0.9$, is compatible but somewhat smaller than the one at PS-191, $R_{\rm observed}/R_{\rm expected} =(2\pm0.5)/(0.7 \pm 0.2)$. The collaboration attributed this to unknown systematic errors in both experiments.

\paragraph{E-734}\label{app:E734} 
E-734 at Brookhaven National Laboratory ran with peak energy $E_\nu^{\rm peak} \sim 1.3$~GeV at a baseline of $\sim 96$~m, and searched for $\nu_\mu \to \nu_e$ transitions~\cite{Ahrens:1984gp}.
The experiment utilized a filter program to remove events not containing a single electromagnetic shower within an angular interval $\theta_e<240$~mrad relative to the beam direction, with the remaining events scanned by physicists to remove events with more than one shower or additional hadronic activity. It is interesting to note those events with one shower and an associated upstream vertex were used as a control sample of photons.
After a cut on the energy, $0.21<E_e\leq5.1$~GeV, $873$ shower events remained. The main backgrounds were identified to be pion production in NC interactions, charged pion production in inelastic CC processes, and those from $\nu_\mu - e$ scattering. Of particular relevance is their cut on the shower energy of $E_e<0.9$~GeV, reducing the sample to $653$ events. 
The final sample contained $418$ events in the energy range $0.9<E_e\leq5.1$~GeV. 

While the experiment saw no excess, we note that most events in our model would not have passed the more stringent cuts. This is mainly due to the larger energies required by the experiment, but also due to the cuts in energy loss, dE/dx, of the shower. Our events would most likely resemble those of the upstream photon-like sample.

\paragraph{E-776}\label{app:E776}
E-776, running with both a narrow- (NBB)~\cite{Blumenfeld:1989qm}  and wide- (WBB)~\cite{Borodovsky:1992pn} band beam of mean energy $1.4$~GeV, searched for $\nu_e$ appearance $1$~km from the target. A fine grained ECAL consisting of $90$ planes of proportional drift tubes interleaved with $1$~in. ($\sim0.25$~ the radiation length) thick concrete absorbers was utilized. A total of $12.8\times10^3$ events were in the full sample, and $1496$ shower events were selected in a scan of the sample. After cuts, which included a requirement that $E_e>600$~MeV, only $55$ events remained. Further cuts on EM shower identification were made, e.g. the number of hits in a cluster and the length of the shower. This left a sample of $38$ events. To eliminate the $\pi^0$ background, the differences in shower profile of pions and electrons were accounted for - the former being wider and more asymmetric. This cut was quoted to have an efficiency of $\sim80\%$ for rejecting $\pi^0$s at $1$~GeV. The final sample contained $17$ electron shower-like events, with the remaining $21$ constituting the $\pi^0$s. Accounting for the probability of pion-electron mis-ID gave $9.6$ events. The observed $17$ events was consistent with the background prediction of $18\pm4.3\text{ (stat.)}\pm3.9\text{ (sys.)}$ events ($9.6\pm3.8\text{ (sys.)}$ from $\pi^0$s and $8.8\pm1.1\text{ (sys.)}$ from $\nu_e$s in the beam), and no excess was reported by the experiment.

Due to the cuts on energy and, in particular, shower profile, a large number of our events would be removed in the analysis, weakening the constraint on our model.

\paragraph{CCFR}\label{app:ccfr}

CCFR searched for production of HNLs with a magnetized toroidal spectrometer-calorimeter, and studied double vertex events from $\nu-N$ interactions.
The sample at CCFR was selected using a neutral current (NC) trigger, whose threshold for energy deposition in the calorimeter was $10$~GeV. To make sure the primary showers were indeed from an NC vertex, it was required that no muons penetrated past the end of the showers. Subsequent cuts selected events with a secondary shower downstream of the first, and further cuts based on kinematical considerations ensured the primary and secondary showers were separated by an angle relative to the beam of $<100$~mrad. The remaining events were categorized by those containing two neutral current vertices (NC/NC), and those with a neutral current vertex followed by a charged current vertex (NC/CC), with the latter being accompanied by a visible muon track in the secondary vertex. For the NC/CC events, cuts on distance between the vertices were made and events with separation $>4\lambda_I$ were selected, where $\lambda_I$ is the nuclear interaction length and $\lambda_I\sim16.8$~cm in Iron. This left $31$ events, which was consistent with the estimated background of $36.8\pm1.7\pm3.4$. For NC/NC events, the backgrounds depended on the shower separation. For long separations, $\gtrsim 14\lambda_I$, the major background was due to random overlay events, events in which independent neutrino interactions appeared correlated. There were negligible contributions from neutral hadron punch-throughs, events in which hadrons created in the initial interaction were able to "punch" through to the end of the shower and interact further downstream. In this region, $9$ events were seen on a background of $3.0\pm0.2\pm0.4$. It is noteworthy that the distributions of some kinematical variables, e.g. hadronic shower energy, for the excess were consistent with those from the overlay background. For those events with separation $<14\lambda_I$, the dominant background was from neutral hadron punch-throughs produced in the initial nuclear interaction. Studies of this background~\cite{deBarbaro:1990sq} suggested there were large degrees of uncertainty, preventing the collaboration presenting results of any excess in this region.

Our model provides an explanation of the excess observed in the NC/NC sample with the prompt decays of $N_6\to N_5 e^+ e^-$, where the $N_5$ travels several meters before decaying to give the secondary shower. At CCFR energies, $\gtrsim 50\%$ of $N_5$s decay within $10$~m. It is also possible to produce the $N_5$s directly in upscattering, giving a signature similar to the above. Alternatively, the first shower can be entirely the hadronic shower from the neutrino interaction vertex with the $N_6$ surviving long enough to decay at the secondary vertex, this is less likely as most $N_6$ will decay within $\lesssim 50$~cm for our BPs.

\bibliography{main}

\end{document}